\documentclass[a4paper,11pt]{article}
\pdfoutput=1 
\usepackage{jheppub} 
\usepackage{epsfig}
\usepackage{caption}
\usepackage{subcaption}
\usepackage{color,soul}
\usepackage{float}
\usepackage{amsfonts}
\usepackage{url}
\usepackage{slashed}
\usepackage{accents}
\usepackage{lipsum}
\usepackage{mathtools}
\usepackage{physics}
\usepackage[normalem]{ulem}
\usepackage{epsfig}
\usepackage[normalem]{ulem}

\hypersetup{
  bookmarks=true,         % show bookmarks bar?
  unicode=false,          % non-Latin characters in Acrobat?s bookmarks
  pdftoolbar=true,        % show Acrobat?s toolbar?
 pdfmenubar=true,        % show Acrobat?s menu?
 pdffitwindow=true,     % window fit to page when opened
 pdfstartview={FitH},    % fits the width of the page to the window
 pdfsubject={Dark Matter},   % subject of the document
 pdfnewwindow=true,      % links in new window
 pdfcreator={RevTeX},
 colorlinks=true,       % false: boxed links; true: colored links
 linkcolor=red,          % color of internal links
 citecolor=blue,        % color of links to bibliography
 filecolor=black,      % color of file links
 urlcolor=blue,           % color of external links
  }

\title{Cannibal dark matter decoupled from standard model: cosmological constraints}
\author[]{Avirup Ghosh,}
\author[]{Sourav Gope}
\author[]{and Satyanarayan Mukhopadhyay}
\affiliation[]{School of Physical Sciences, Indian Association for the Cultivation of Science, 2A and 2B Raja S.C. Mullick Road, Kolkata 700 032}

\emailAdd{spsag2510@iacs.res.in}
\emailAdd{intsg5@iacs.res.in}
\emailAdd{tpsnm@iacs.res.in}

\abstract{An internally thermalized dark matter (DM) with only gravitational interaction with the standard model (SM) particles at low temperatures, may undergo number-changing self-scatterings in the early Universe, eventually freezing out to the observed DM abundance. If these reactions, such as a $3 \rightarrow 2$ process, take place when the DM is non-relativistic, DM cannibalizes itself to cool much slower than standard non-relativistic matter during the cannibal phase. As shown in earlier studies, if the cannibal phase takes place during the matter-dominated epoch, there are very strong constraints from structure formation. Considering scenarios in which the cannibal phase freezes out in the radiation-dominated epoch instead, we show that cannibal DM decoupled from the SM can be viable, consistent with all present cosmological constraints. To this end, we solve the coupled evolution equations of the DM temperature and density, and determine its abundance for different DM self-couplings. We then evaluate the constraints on these parameters from the cosmic-microwave background power spectrum, the big-bang nucleosynthesis limits on the relativistic degrees of freedom, the Lyman-$\alpha$ limits on the DM free-streaming length and the theoretical upper bound on the $3 \rightarrow 2$ annihilation rate from $S-$matrix unitarity. We find that depending upon the DM self-couplings, a scalar cannibal DM with mass in the range of around 80 eV to 700 TeV can make up the observed DM density and satisfy all the constraints, when the initial DM temperature ($T_{\rm DM}$) is lower than the SM one ($T_{\rm SM}$), with $T_{\rm SM}/9100 \lesssim T_{\rm DM} \lesssim \,T_{\rm SM}/1.1$.}

\begin{document} 
\maketitle
\flushbottom  
\section{Introduction}
\label{sec:sec1}
The possibility of dark matter (DM) undergoing number-changing self-interactions was first introduced by Dolgov in Refs.~\cite{Dolgov:1980uu, Dolgov:2017ujf}, in the context of DM being a glueball of a strongly interacting hidden gauge theory. This was perceived as a mechanism by which a self-thermalized DM, with only gravitational interaction with the standard model (SM), may deplete its number density in order not to overclose the Universe. The same idea was also proposed by Carlson, Machacek and Hall in Ref.~\cite{Carlson:1992fn} with a DM candidate which behaves, in the context of structure formation in the Universe, differently from hot DM or cold DM (CDM), thereby accommodating possible departures in the matter power spectrum from CDM predictions. 

In such scenarios, the DM elastic self-scatterings need to be at a rate necessary to maintain an internally thermalized DM with its own temperature. As long as the number-changing reactions, such as a $3 {~\rm DM}\rightarrow 2{~\rm DM}$ process, are in chemical equilibrium, the chemical potential of the DM vanishes. These self-scattering reactions are exothermic in nature -- if they take place when the DM is non-relativistic (NR) with $T_\chi < m_\chi$, $T_\chi$ and $m_\chi$ being the temperature and the mass of the DM respectively, the DM cannibalizes itself to keep warm, with its temperature falling as $T_\chi \sim 1/\log{a}$, while its energy density falls as $\rho_\chi \sim 1/(a^3 \log{a})$, $a(t)$ being the scale-factor in the standard Friedmann-Robertson-Walker (FRW) metric. This is in contrast to the $1/a^2$ fall of the ordinary NR matter temperature, or the $1/a$ fall of the temperature of a relativistic species. As observed in Refs.~\cite{Carlson:1992fn, Machacek:1994vg, deLaix:1995vi}, this very different evolution compared to ordinary NR matter may significantly impact large-scale structure formation of the Universe. 

The earlier studies in Refs.~\cite{Carlson:1992fn, Machacek:1994vg, deLaix:1995vi}, and the more recent detailed analyses in Refs.~\cite{Buen-Abad:2018mas, Heimersheim:2020aoc} on cannibal DM were primarily motivated to explore if possible departures in the matter power spectrum from the CDM predictions can be addressed using cannibal DM. As emphasized in all of these works, in order for the cannibal phase to have a considerable impact on structure formation, it must take place during the matter-dominated epoch of the Universe, and therefore the cannibal freeze-out happens after matter-radiation equality. Such a possibility is, of course, strongly constrained by the precise determination of the matter power spectrum from the cosmic microwave background (CMB) data, though viable regions accommodating small-scale anomalies exist~\cite{Buen-Abad:2018mas, Heimersheim:2020aoc}.

In this paper, we study instead in detail the possibility of cannibal freeze-out during the radiation dominated epoch of the Universe. Therefore, for this study, we shall assume that the observed matter power-spectrum is well described by the CDM cosmology. The main objective of our study is to determine if a cannibal DM, with only gravitational coupling to the SM sector at low temperatures, can be a plausible particle DM candidate making up the observed DM density, at the same time satisfying all other cosmological constraints. 

In order to address this question, we have accurately modelled the evolution of the DM temperature and density by solving the coupled system of evolution equations, and determined the viable parameter region consistent with the relic density requirement, details of which are presented in Sec.~\ref{sec:sec2}. We have then evaluated the observational constraints from the cosmic microwave background power spectrum, big-bang nucleosynthesis (BBN) constraints on the relativistic degrees of freedom, the Lyman-$\alpha$ constraints on the DM free-streaming length and the theoretical upper bound on the $3 \rightarrow 2$ annihilation rate from $S-$matrix unitarity. Our results on the allowed region in the DM temperature-mass plane in which the DM relic density can describe the observed abundance, along with the various cosmological and theoretical constraints are presented in Sec.~\ref{sec:sec3}. We summarize our findings in  Sec.~\ref{sec:sec4}, and provide further computational details on performing collision integrals in Appendix~\ref{Appendix}.

This study should fill a gap in the literature where a clear picture of the allowed region of the DM temperature and mass values in this scenario is missing. In particular, we emphasize that, such a scenario is very much allowed by the structure formation constraints, even when cannibal DM cannot dissipate the heat generated during the cannibal phase to the SM sector, as long as the DM is sufficiently cold compared to the SM to begin with, but not so cold that chemical equilibrium of the $3 \rightarrow 2$ process is not achieved. Thus, the scenario considered in our study is very different from the so-called SIMP DM, in which the DM is kinetically coupled to and thermalized with the SM sector~\cite{Hochberg:2014dra}. For studies on other interesting facets of cannibal DM, we refer the reader to Refs.~\cite{Pappadopulo:2016pkp, Farina:2016llk, Erickcek:2020wzd, Erickcek:2021fsu}.

\section{Evolution of cannibal dark matter temperature and density}
\label{sec:sec2}
An internally thermalized cannibal DM $\chi$ undergoing $3\chi \rightarrow 2\chi$ reactions in chemical equilibrium, is described by its mass $m_\chi$, temperature $T_\chi$ and the rates of the number changing $3\chi \rightarrow 2\chi$ and number conserving $2\chi \rightarrow 2\chi$ reactions. The latter rate is important in ensuring that the DM remains internally thermalized, at least throughout the phase the $3\chi \rightarrow 2\chi$ reactions are taking place. Within such a simplistic setup, the DM temperature and mass are free parameters, with the temperature being determined by the initial conditions in the very early Universe, presumably set at the reheating epoch in the context of inflationary cosmology. The reaction rates are determined in terms of the DM self-couplings, as will be illustrated in the following. 

The basic equation determining the evolution of the DM phase-space density $f_\chi(\textbf{p},t)$ is the Boltzmann kinetic equation given by
\begin{equation} 
\frac{\partial\,f_\chi(\textbf{p},t)}{\partial\,t} - H\textbf{p}.\nabla_{\textbf{p}} f_\chi(\textbf{p},t) = C[f_{\chi}],
\label{Eq:Boltz1}
\end{equation}
where $C[f_{\chi}]$ encodes the collision terms for all the elastic and inelastic processes affecting the distribution function $f_\chi(\textbf{p},t)$, and $H$ is the Hubble expansion rate. Integrating Eqn.~\ref{Eq:Boltz1} over the momenta $\textbf{p}$ yields the equation for the DM number density $n_\chi(t)$ as
\begin{equation} 
\frac{dn_{\chi}(t)}{dt}+3Hn_{\chi}(t) = g_{\chi}\int \frac{d^3\textbf{p}}{(2\pi)^3}C[f_{\chi}] \equiv C_0,
\label{Eq:Boltz_number}
\end{equation}
where the number density for a particle with $g_\chi$ number of internal degrees of freedom is defined by 
\begin{equation} 
n_{\chi}=g_{\chi}\int \frac{d^3\textbf{p}}{(2\pi)^3}f_{\chi}(\textbf{p},t).
\end{equation}

We can define the temperature of a species as the average of $|\textbf{p}|^2/3E$ over its distribution function as follows~\cite{Bringmann:2006mu}
\begin{equation}
\label{Eq:Temperature_Defn}
T_{\chi} \equiv \frac{g_{\chi}}{n_{\chi}}\int \frac{d^3\textbf{p}}{(2\pi)^3} \frac{|\textbf{p}|^2}{3E} f_{\chi}(\textbf{p},t).
\end{equation}
As it can be readily verified with explicit computation, this definition is an identity for both a relativistic and a non-relativistic species with an equilibrium distribution. On taking the moment of the Boltzmann equation~\ref{Eq:Boltz1} with $|\textbf{p}|^2/3E$, we arrive at the evolution equation for the DM temperature as 
\begin{equation}
\frac{dT_\chi}{dt}+2 H T_\chi +\frac{T_\chi}{n_\chi} \left (\frac{dn_{\chi}}{dt}+3Hn_{\chi} \right) - \frac{H}{3} \left\langle \frac{|\textbf{p}|^4}{E^3} \right\rangle = \frac{C_2}{n_\chi},
\label{Eq:temp_evolve}
\end{equation}
where, $C_2$ stands for the following expression
\begin{equation}
C_2 = \int \frac{d^3\textbf{p}}{(2\pi)^3} \frac{|\textbf{p}|^2}{3E} C[f_{\chi}].
\end{equation}
Here, $\left\langle |\textbf{p}|^4/E^3 \right\rangle$ denotes the average of $|\textbf{p}|^4/E^3$ over the DM phase-space distribution.

In order to obtain explicit expressions for the collision terms appearing in Eqs.~\ref{Eq:Boltz_number} and~\ref{Eq:temp_evolve}, we need to define the relevant processes. For the cannibal DM scenario under study, the simplest example would be to take a real scalar DM $\chi$, with $\chi \chi \rightarrow \chi \chi$ elastic scattering, and $3 \chi  \rightarrow 2 \chi$ inelastic scattering reactions. With these, the collision integral for the DM number density equation receives contributions only from the inelastic number changing reactions, and is given by
\begin{align}
C_0 = \frac{1}{3!2!}\int d\Pi_1 d\Pi_2 d\Pi_3 d\Pi_4 d\Pi_5 (2\pi)^4 \delta^4(p_1+p_2+p_3-p_4-p_5)|\mathcal{M}|^2_{3\chi\rightarrow2\chi} \nonumber\\
\big[f_{\chi}(E_4)f_{\chi}(E_5)-f_{\chi}(E_1)f_{\chi}(E_2)f_{\chi}(E_3)\big]
\label{Eq:Czero_1}
\end{align} 
where $d\Pi_j\equiv\frac{d^3\textbf{p}_j}{(2\pi)^3 2E_j}$, and $|\mathcal{M}|^2_{3\chi\rightarrow2\chi}$ denotes the matrix element squared for the $3\chi\rightarrow2\chi$ process. Due to the isotropy of the background cosmology, the distribution functions depend only on $|\textbf{p}|$, and hence can be expressed as a function of $E$. We can express the collision integral in Eq.~\ref{Eq:Czero_1} in terms of the thermally averaged reaction rate as
\begin{equation} 
\label{Eq:C0_defn}
C_0 = \langle \sigma v^2 \rangle_{3\chi\rightarrow2\chi} \left( n_{\chi}^2 n_{\chi}^{\rm eq}-n_{\chi}^3 \right)
\end{equation}
where,  $n_{\chi}^{\rm eq}$ is the equilibrium number density of $\chi$, and $\langle \sigma v^2 \rangle_{3\chi\rightarrow2\chi}$ is defined by
\begin{align}
 \langle \sigma v^2 \rangle_{3\chi\rightarrow2\chi} = \frac{1}{3!}\frac{1}{(n_{\chi}^{\rm eq})^3}\int d\Pi_1 d\Pi_2 d\Pi_3  (2E_1 2E_2 2E_3) (\sigma v^2)   
 f_{\chi}^{\rm eq}(E_1)f_{\chi}^{\rm eq}(E_2)f_{\chi}^{\rm eq}(E_3),
 \label{Eq:sigmav2}
\end{align} 
with $\sigma v^2$ given by
\begin{align}
 \sigma v^2  = \frac{1}{2!} \frac{1}{2E_1 2E_2 2E_3}\int   d\Pi_4 d\Pi_5  (2\pi)^4 \delta^4(p_1+p_2+p_3-p_4-p_5)  |\mathcal{M}|^2_{3\chi\rightarrow2\chi}.
 \label{Eq:sigmav2_noavg}
\end{align} 

Let us now consider the collision term in the temperature evolution equation~\ref{Eq:temp_evolve} due to the $\chi(\textbf{p}_1)+\chi(\textbf{p}_2)+\chi(\textbf{p}_3)\rightarrow \chi(\textbf{p}_4)+\chi(\textbf{p}_5)$ process. When any one of the identical particles with momentum $\textbf{p}_1, \textbf{p}_2$ or $\textbf{p}_3$ is in the initial state, these particles lose energy due to the collision, while the identical particles with momenta $\textbf{p}_4$ or $\textbf{p}_5$ gain energy due to the same process, and vice versa. Therefore, the collision term can be written, by symmetrizing with respect to the incoming and outgoing particle momenta, as~\cite{Buen-Abad:2018mas}
\begin{align} 
C_2 = \frac{1}{3!2!}\int d\Pi_1 d\Pi_2 d\Pi_3 d\Pi_4  d\Pi_5 \Big( \frac{|\textbf{p}_1|^2}{3E_1} +\frac{|\textbf{p}_2|^2}{3E_2}+\frac{|\textbf{p}_3|^2}{3E_3}-\frac{|\textbf{p}_4|^2}{3E_4}-\frac{|\textbf{p}_5|^2}{3E_5}\Big) \nonumber \\
(2\pi)^4\delta^4(p_1+p_2+p_3-p_4-p_5)|\mathcal{M}|^2_{\textbf{p}_1,\textbf{p}_2,\textbf{p}_3 \rightarrow \textbf{p}_4,\textbf{p}_5}  \big(f_{\chi}(E_4)f_{\chi}(E_5)-f_{\chi}(E_1)f_{\chi}(E_2)f_{\chi}(E_3)\big)
\label{Eq:C2}
\end{align}
which can be further reduced to the following form
\begin{equation}
C_2 = \left \langle \sigma v^2 .\Big( \frac{|\textbf{p}_1|^2}{3E_1} +\frac{|\textbf{p}_2|^2}{3E_2}+\frac{|\textbf{p}_3|^2}{3E_3}-\frac{|\textbf{p}_4|^2}{3E_4}-\frac{|\textbf{p}_5|^2}{3E_5}\Big) \right \rangle_{3\chi\rightarrow2\chi} \left( n_{\chi}^2n_{\chi}^{\rm eq}-n_{\chi}^3 \right),
\label{Eq:C2_reduced}
\end{equation}
where the thermal average of the quantity in angle brackets $\langle ... \rangle$ is defined in a similar way to Eq.~\ref{Eq:sigmav2}, with $\sigma v^2$ given by Eq.~\ref{Eq:sigmav2_noavg}.

In order to present our numerical results, we need to define an appropriate matrix element squared for the $3\chi \rightarrow 2\chi$ process. For this purpose, we adopt a simple low-energy toy model for cannibal DM, with a real scalar field $\chi$ interacting through the following interaction Lagrangian density:~\footnote{The stability of the real scalar cannibal DM, as well as the absence of its interactions with the SM can follow from its ultra-violet dynamics. For example, Refs.~\cite{Dolgov:1980uu, Dolgov:2017ujf, Carlson:1992fn, Buen-Abad:2018mas} considered $\chi$ as the glueball of a hidden non-abelian gauge theory. Instead of a real scalar field, one can also consider a very similar low-energy toy model with a complex scalar field stabilized by a $Z_3$ symmetry, including again both the cubic and quartic scalar interaction terms. All our subsequent discussions will hold in this case as well, with a change of the number of degrees of freedom for the DM to $2$, and with the inclusion of appropriate symmetry factors. In this case, the relic density will be given by the sum of the DM particle and anti-particle contributions, assuming no asymmetry is generated.}
\begin{equation}
\mathcal{L}_{\rm int} = -\frac{\mu}{3!}\chi^3 -\frac{\lambda}{4!}\chi^4.
\label{Eq:Lag}
\end{equation}
Evaluating the matrix element squared for the process $\chi(\textbf{p}_1)+\chi(\textbf{p}_2)+\chi(\textbf{p}_3)\rightarrow \chi(\textbf{p}_4)+\chi(\textbf{p}_5)$ with the interaction Lagrangian in Eq.~\ref{Eq:Lag}, one obtains in the non-relativistic approximation:
		
\begin{equation}
|\mathcal{M}|^2 \simeq \frac{25\,\mu^2}{64\,m^4_\chi}\left(3\lambda - \frac{\mu^2}{m^2_\chi} \right)^2.
\end{equation}
The details of the matrix element computation and the relevant Feynman diagrams are shown in Appendix~\ref{Appendix}.	  

\begin{figure} [htb!]
	\begin{center} 
		\includegraphics[scale=0.18]{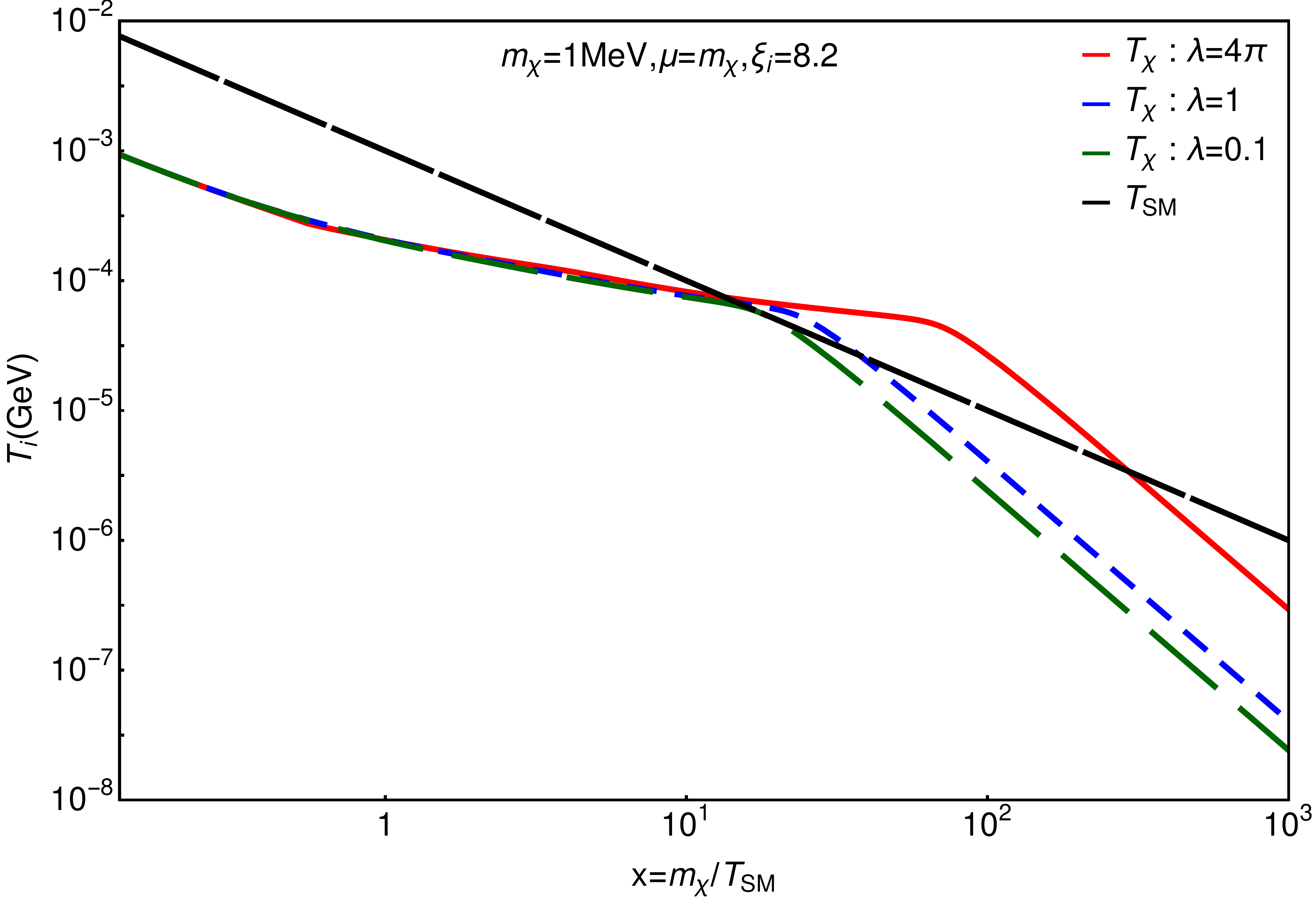} \hspace{0.0cm}
		\includegraphics[scale=0.215]{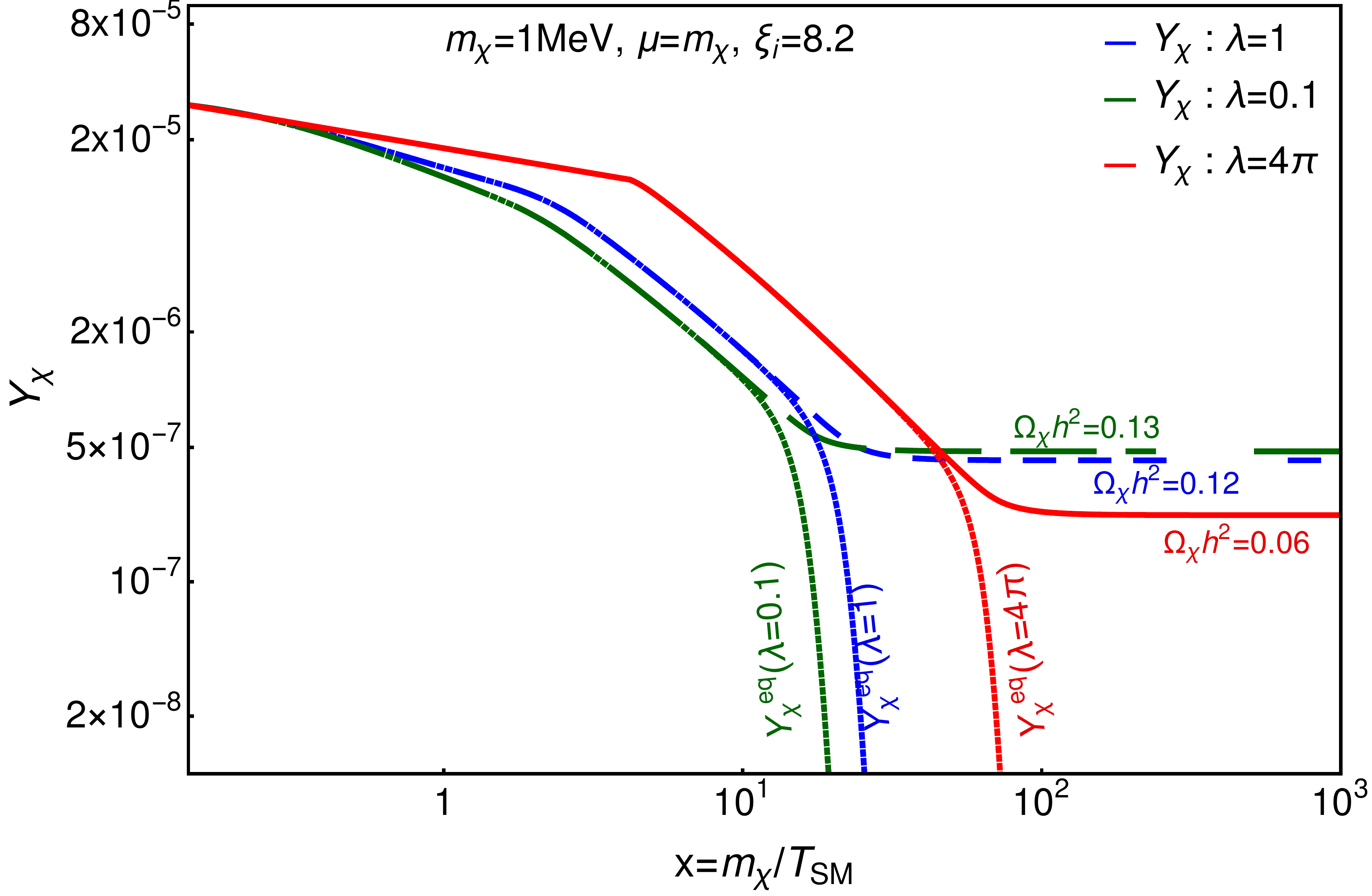}
	\end{center}
	\caption{\small{\it{{\bf Left panel:} Evolution of the DM temperature $T_\chi$ as a function of the dimensionless variable $x=m_\chi/T_{\rm SM}$, for different values of the DM quartic coupling $\lambda=0.1,1$ and $4\pi$, with the other parameters fixed as the DM mass $m_\chi=1 {~\rm MeV}$, the trilinear DM coupling $\mu=m_\chi$, and the initial ratio of the SM and DM temperatures $\xi_i= 8.2$. For comparison, the SM temperature $T_{\rm SM}$ is also shown. {\bf Right panel:} The DM yield $Y_\chi = n_\chi/s$, where $s$ is the entropy density of the SM bath, as a function of $x$, for the same set of parameters as in the left panel. For comparison, the equilibrium distribution functions $Y_\chi^{\rm eq} (x)$ are also shown. The resulting relic abundance $\Omega_\chi h^2$  of the cannibal DM are  indicated in each case in the right panel. See text for details.} } }
	\label{Fig:Temp_evol}
\end{figure}

For the following discussion, it will be convenient to define the ratio of the SM and the DM temperatures at a particular epoch as 
\begin{equation}
\xi(a) = T_{\rm SM}(a) / T_{\chi}(a),
\end{equation}
where $a(t)$ is the scale factor at that epoch. The temperature ratio $\xi(a)$ remains constant at its initial value $\xi_i$ until the epoch with scale factor $a_{\rm NR}$ when the DM becomes non-relativistic, with $T_\chi(a_{\rm NR})/m_\chi \sim 1$. This is because, until this epoch, both the SM and the DM temperatures fall as $1/a(t)$, characteristic of relativistic species. 

Both the multi-dimensional phase-space integrals in Eqs.~\ref{Eq:sigmav2} and~\ref{Eq:C2_reduced} can be further simplified by considering the kinematics of the $3\chi \rightarrow 2\chi$ process, as detailed in the Appendix~\ref{Appendix}. The resulting integrals are performed using the Monte-Carlo integration method implemented in $\texttt{CUBA}$~\cite{Hahn:2004fe}. With the obtained collision integrals as inputs to the number density and temperature evolution Eqs.~\ref{Eq:Boltz_number} and~\ref{Eq:temp_evolve}, we solve these coupled differential equations using the $\texttt{FORTRAN}$ library $\texttt{DDASSL}$~\cite{DDASSL}. 

\begin{figure} [htb!]
	\begin{center} 
		\includegraphics[scale=0.22]{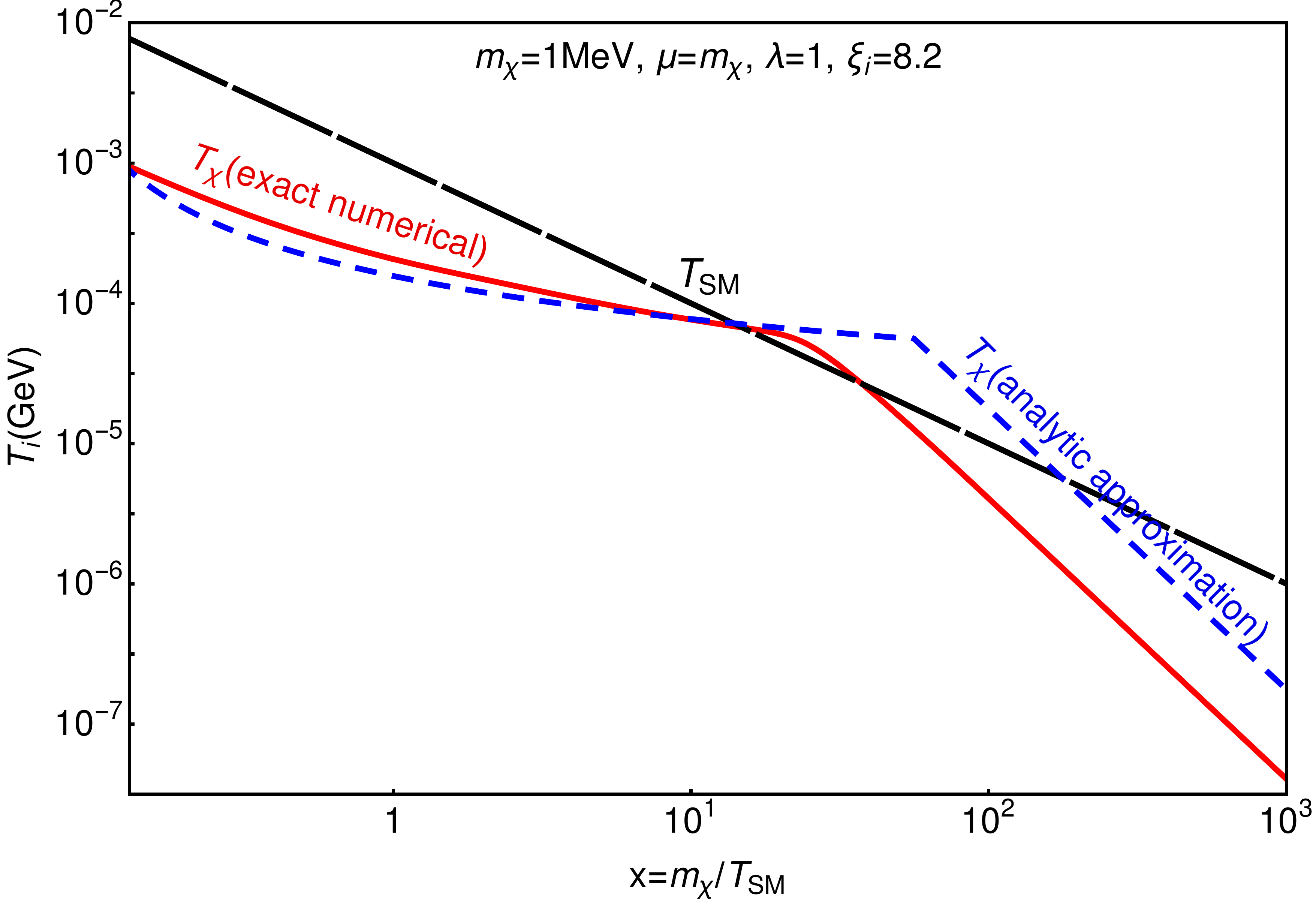} \hspace{0.0cm}
	\end{center}
	\caption{\small{\it{Comparison of the evolution of the DM temperature $T_\chi$ as a function of the dimensionless variable $x=m_\chi/T_{\rm SM}$, obtained with the exact numerical solution of the evolution Eqs.~\ref{Eq:Boltz_number} and~\ref{Eq:temp_evolve}, and the approximate analytic solution obtained using entropy conservation in the DM sector along with the freeze-out approximation. We have set the quartic coupling $\lambda=1$, with all other parameters same as in Fig.~\ref{Fig:Temp_evol}. } } }
	\label{Fig:number_evol}
\end{figure}

We show the numerical solutions to Eqs.~\ref{Eq:Boltz_number} and~\ref{Eq:temp_evolve} in Figs.~\ref{Fig:Temp_evol} and ~\ref{Fig:number_evol}. In the left panel of Fig.~\ref{Fig:Temp_evol}, we show the DM temperature $T_\chi$ as a function of the dimensionless variable $x=m_\chi/T_{\rm SM}$, the latter being used as a measure for the flow of time. The results are shown for three different values of the DM quartic coupling $\lambda=0.1,1$ and $4\pi$, with green long-dashed, blue dashed and red solid lines, respectively. In each case, the other parameters have been fixed as follows: the DM mass $m_\chi=1 {~\rm MeV}$, the trilinear DM coupling $\mu=m_\chi$, and the initial ratio of the SM and DM temperatures $\xi_i=8.2$. The value of $\xi_i$ is chosen such that the observed DM abundance $\Omega_\chi h^2=0.12$~\cite{Planck:2018vyg} is given by the relic density of the cannibal DM $\chi$ with $\lambda=1$. 
For comparison, we also show the SM temperature by the diagonal black dashed line. As we can see from this figure, until the freeze-out of the cannibal process, the DM temperature falls as $T_\chi \propto 1/\log(a)$, much slower than the SM temperature. After the cannibal process freezes out, the DM temperature falls faster than the SM temperature as $T_\chi \propto 1/a^2$, as characteristic of ordinary NR matter. 

In order to understand the decoupling point of the cannibal process better, in the right panel of Fig.~\ref{Fig:Temp_evol} we show the DM yield $Y_\chi = n_\chi/s$, where $s$ is the entropy density of the SM bath, as a function of $x$, for the same set of parameters used as in the left panel. For comparison, the equilibrium distribution functions $Y_\chi^{\rm eq}$ in each case are also shown with dotted lines. Since $Y_\chi^{\rm eq}$ here is shown as a function of $x$, and not $x^\prime = m_\chi/T_\chi$, and the evolution of $T_\chi$ as a function of $x$ depends upon the coupling $\lambda$, we obtain a different $Y_\chi^{\rm eq} (x)$ for different values of the quartic coupling. The value of $x$ at which the DM yield freezes, and hence the $3\chi \rightarrow 2\chi$ cannibal process stops taking place can be read off from this figure, which are the same points in which the turnover from the logarithmic to the quadratic fall in temperature can be observed in the left panel. We have also indicated the values of the DM relic abundance $\Omega_\chi h^2$ in the right panel of Fig.~\ref{Fig:Temp_evol}. One observes that the relic density evolves very slowly as a function of the coupling $\lambda$, and a factor of $125$ increase in the coupling only leads to a factor of 2.2 decrease in the abundance. On the other hand, for the same variation of the quartic coupling, the DM temperature at large values of $x$ may vary by upto one order of magnitude, thereby affecting the resulting cosmological constraints  
significantly.

Although the qualitative behaviour of the DM temperature evolution as $T_\chi \propto 1/\log(a)$ in the cannibal phase can be obtained using the separate conservation of the DM entropy, the precise point at which the turnover from the cannibal phase to the NR phase with $T_\chi \propto 1/a^2$ occurs can only be obtained approximately using the freeze-out condition of $ n_\chi^2 (a_{\rm FO}) \langle \sigma v^2 \rangle (a_{\rm FO}) = H(a_{\rm FO}) $. The freeze-out point obtained through the exact numerical solution of Eqs.~\ref{Eq:Boltz_number} and~\ref{Eq:temp_evolve} differs considerably from this estimate, and therefore the value of the DM temperature at large $x$ can be modified 
nearly by an order of magnitude. This is illustrated in Fig.~\ref{Fig:number_evol}, where we compare the exact numerical solution for the evolution of $T_\chi$ (red solid line) and the analytical approximation discussed above (blue dashed line) for the $\lambda=1$ scenario, with all other parameters fixed at the same value as in Fig.~\ref{Fig:Temp_evol}.

\section{Cosmological constraints: allowed range of DM temperature and mass}
\label{sec:sec3}
\begin{figure} [htb!]
\begin{center} 
\includegraphics[scale=0.36]{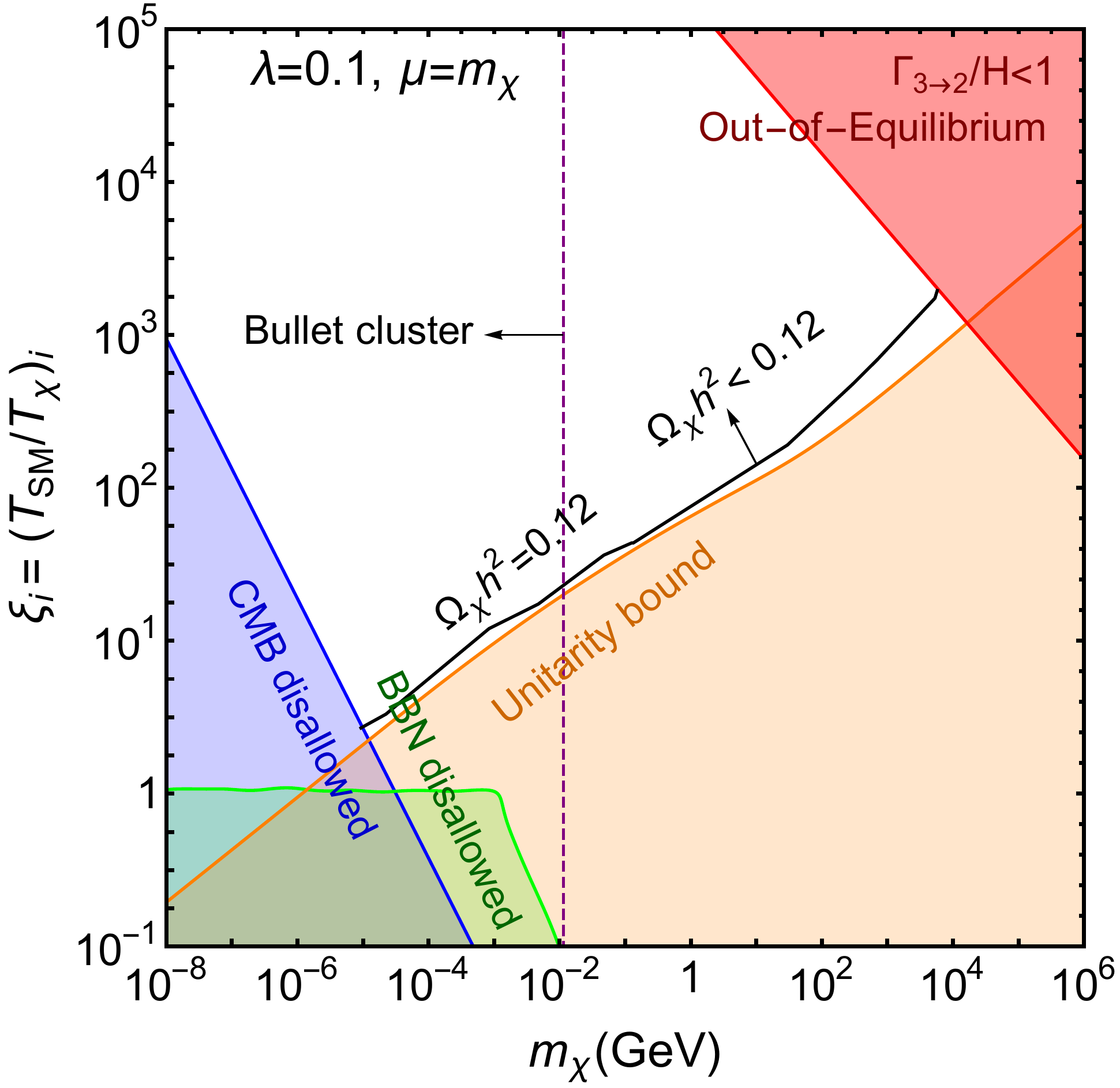} \hspace{0.0cm}
\includegraphics[scale=0.36]{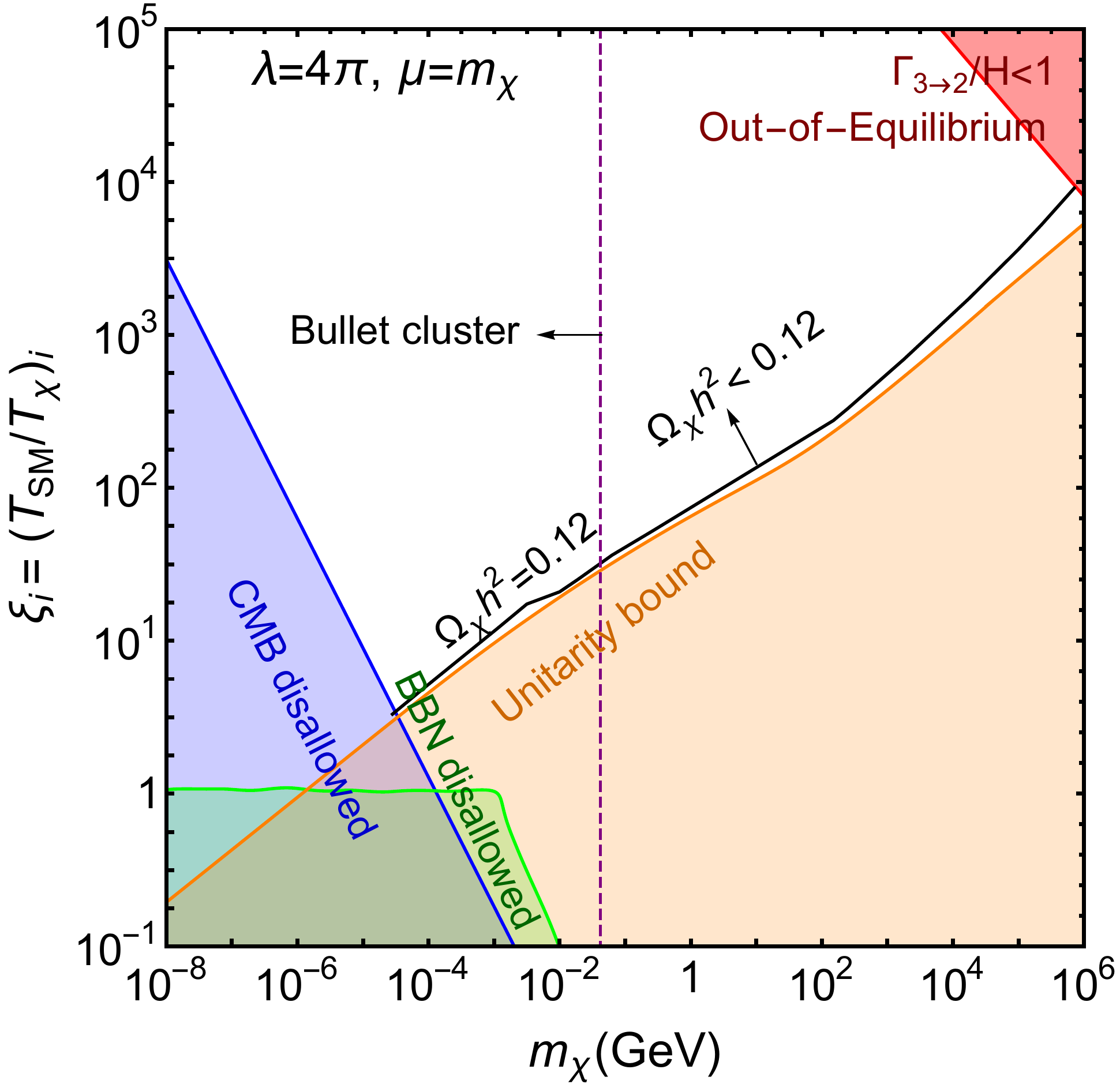}
\end{center}
\caption{\small{\it{Constraints on the dark matter mass $m_\chi$ and the initial SM to DM temperature ratio $\xi_i = (T_{\rm SM}/T_\chi)_i$ plane from CMB and BBN data, from the requirement of chemical equilibrium for the $3\chi \rightarrow 2\chi$ process, and from the unitarity upper bound on the DM inelastic scattering rate. Also shown is the contour along which the cannibal DM reproduces the observed DM abundance $\Omega_\chi h^2 = 0.12$. The results are shown for the DM quartic coupling $\lambda=0.1$ (left panel) and $\lambda=4\pi$ (right panel), with the trilinear coupling fixed as $\mu=m_\chi$. The Bullet Cluster constraints on DM elastic self-scatterings,  $\sigma_{2\chi \rightarrow 2\chi}/m_\chi < 1 {~\rm cm}^2/{\rm g}$, are shown by the vertical 
dashed lines, with the region in the direction of the arrow being disfavoured, subject to large astrophysical 
uncertainties (for details, see Sec.~\ref{sec:LyalphaBul}). }}}
\label{Fig:Cosmo_constraint1}
\end{figure}
As mentioned earlier, the properties of cannibal DM are determined by its initial temperature $T_\chi$, its mass $m_\chi$ and the reactions rates for the elastic $2\chi \rightarrow 2\chi$ and the inelastic $3\chi \rightarrow 2 \chi$ processes. In the simple toy model under consideration with a real scalar DM $\chi$ discussed in the previous section, the reaction rates are determined by the scalar trilinear coupling $\mu$ and the quartic coupling $\lambda$. We observed from Figs.~\ref{Fig:Temp_evol} and ~\ref{Fig:number_evol} that with a large variation in these couplings, while the relic abundance of $\chi$ is modified only by a small factor, the DM temperature at late times is modified significantly. As the DM temperature controls the cosmological constraints on cannibal DM, we expect these constraints to be modified substantially while varying the self-couplings. 

\begin{figure} [htb!]
\begin{center} 
\includegraphics[scale=0.36]{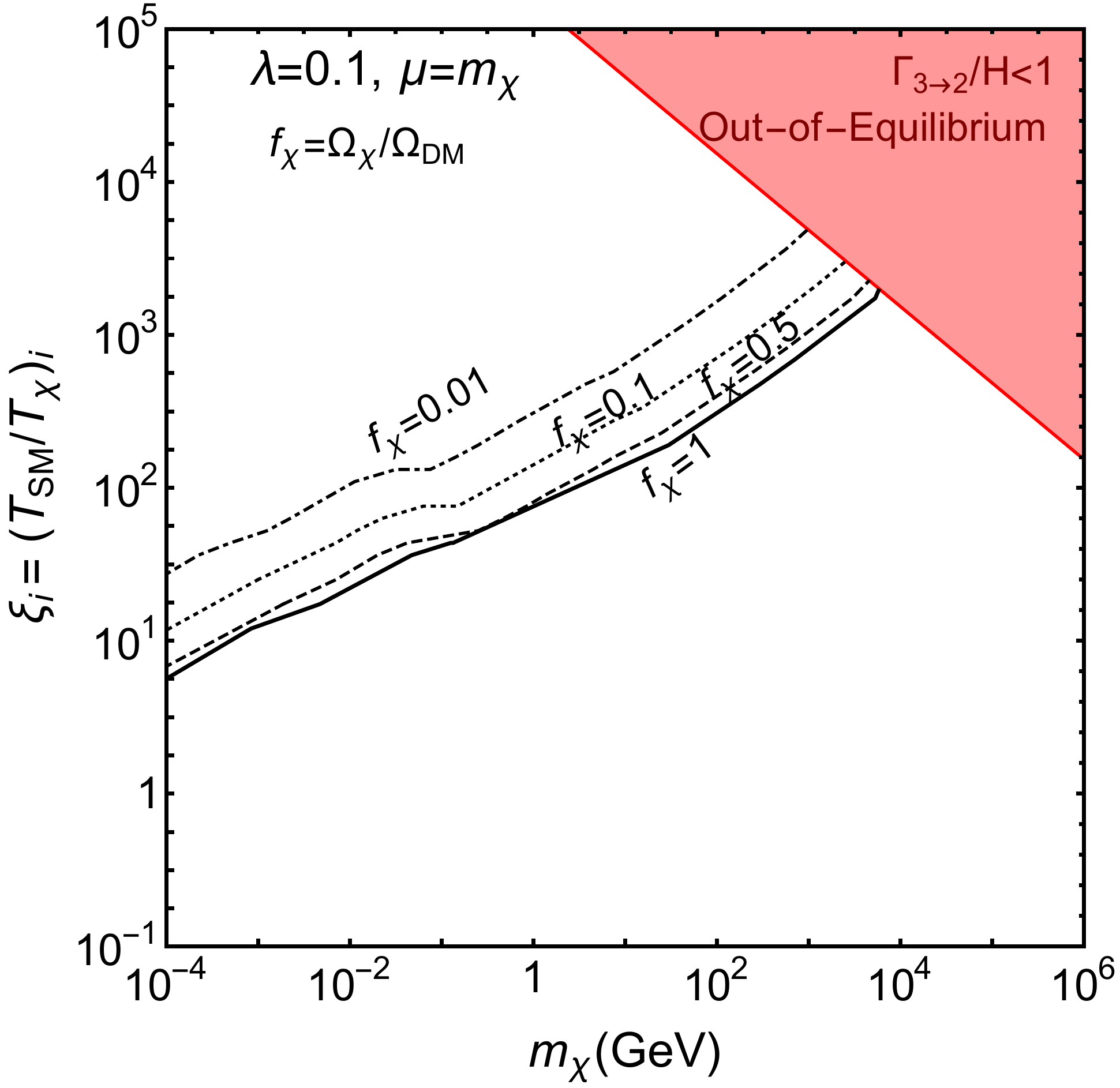} \hspace{0.0cm}
\end{center}
\caption{\small{\it{The DM fraction $f_\chi = \Omega_\chi / \Omega_{\rm DM}$ in the parameter space of Fig.~\ref{Fig:Cosmo_constraint1}, with the DM quartic coupling $\lambda=0.1$ and trilinear coupling $\mu=m_\chi$.}}}
\label{Fig:DM_fraction}
\end{figure}

In this section, we shall focus on the different constraints on the DM temperature and mass coming from the CMB, BBN and Lyman-$\alpha$ data. We shall also consider the requirement of the chemical equilibrium of the $3\chi \rightarrow 2\chi$ process, which is the starting premise of this study. Finally, we study the implications of $S-$matrix unitarity, which sets an upper limit on the total annihilation rate of the $3\chi \rightarrow 2\chi$ process through the optical theorem. In Fig.~\ref{Fig:Cosmo_constraint1}, we show the resulting constraints on the dark matter mass $m_\chi$ and the initial SM to DM temperature ratio $\xi_i = (T_{\rm SM}/T_\chi)_i$ plane. Also shown is the contour along which the cannibal DM reproduces the observed DM abundance $\Omega_\chi h^2 = 0.12$. The results are shown for the DM quartic coupling $\lambda=0.1$ (left panel) and $\lambda=4\pi$ (right panel), with the trilinear coupling fixed as $\mu=m_\chi$. For a given value of the DM mass, while $\Omega_\chi h^2 = 0.12$ is obtained for a particular $\xi_i$ value, higher values of $\xi_i$ lead to scenarios in which the cannibal DM makes up only a fraction of the observed density, as indicated in Fig.~\ref{Fig:Cosmo_constraint1}. In this region, the DM fraction $f_\chi = \Omega_\chi / \Omega_{\rm DM}$ drops rapidly, as illustrated in Fig.~\ref{Fig:DM_fraction}.

We discuss the details of each constraint mentioned above in the following.

\subsection{Chemical equilibrium and freeze-out of the cannibal process}
To begin with, we are interested in the scenario in which the DM relic density is determined by the freeze-out of the $3\chi \rightarrow 2\chi$ cannibal process, where the freeze-out takes place when the DM is non-relativistic. Therefore, we shall work in the region of parameter space where the $3\chi \rightarrow 2\chi$ process is in chemical equilibrium at $x^\prime = m_\chi/T_\chi = 1$. This is ensured by imposing the following condition
\begin{equation}
n^2_{\chi,{\rm eq}} (x^\prime=1) \langle \sigma v^2 \rangle_{3\chi \rightarrow 2\chi} (x^\prime=1) > H (x=1/\xi_i),
\label{Eq:chemical_eq}
\end{equation}
with $x = m_\chi/T_{\rm SM}$, and as defined earlier, $\xi_i = T_{\rm SM}/T_\chi$ is the initial value of the temperature ratio. The region 
where this chemical equilibrium condition Eq.~\ref{Eq:chemical_eq} in the non-relativistic regime of 
the DM is not satisfied is shown by the red shaded area on the top right corner of 
Fig.~\ref{Fig:Cosmo_constraint1}. With a fixed value of DM mass and with $x^\prime=1$, for higher 
values of $\xi_i$, the SM temperature is higher, as $T_{\rm SM} \propto \xi_i$ for a fixed $T_\chi$. 
This in turn implies that the Hubble expansion rate $H (x=1/\xi_i)$ increases with increasing 
$\xi_i$, while the $3\chi \rightarrow 2\chi$ reaction rate $\Gamma_{3 \rightarrow 2}(x^\prime=1) = 
n^2_{\chi,{\rm eq}} (x^\prime=1) \langle \sigma v^2 \rangle_{3\chi \rightarrow 2\chi} (x^\prime=1)$ 
remains the same. Thus we see that for a fixed DM mass, higher $\xi_i$ values do not lead to chemical 
equilibrium in the non-relativistic regime. Furthermore, with the increase in $m_\chi$, the 
annihilation rate $\langle \sigma v^2 \rangle_{3\chi \rightarrow 2\chi}$ decreases as $1/m_\chi^5$, 
for the trilinear coupling $\mu \sim \mathcal{O}(m_\chi)$, in the model described by the interaction 
Lagrangian in Eq.~\ref{Eq:Lag}, and is approximately given by~\cite{Erickcek:2021fsu}

\begin{equation}
\langle \sigma v^2 \rangle_{3\chi \rightarrow 2\chi} \simeq \frac{25\sqrt{5} \mu^2}{147456 \pi m_\chi^7} \left(3\lambda - \frac{\mu^2}{m^2_\chi} \right)^2 .
\label{Eq:sigmav_32}
\end{equation}
The computational details for the above reaction rate are shown in Appendix~\ref{Appendix}.
This reduced reaction rate leads to earlier decoupling, and hence no chemical equilibrium in the higher mass and higher $\xi_i$ region is obtained when the DM is non-relativistic. If the quartic coupling $\lambda$ is increased, one of course obtains a smaller region excluded by the chemical equilibrium criterion. For illustration, we have shown the results in Fig.~\ref{Fig:Cosmo_constraint1} for two values of the quartic coupling, with $\lambda=0.1$ (left panel) and $\lambda=4\pi$ (right panel), keeping $\mu=m_\chi$ fixed.

While the exact value of the scaled DM-temperature $x^\prime$ at the freeze-out of the $3\chi \rightarrow 2\chi$ process, $x^\prime_F$, is determined by solving the coupled evolution equations~\ref{Eq:Boltz_number} and Eq.~\ref{Eq:temp_evolve}, we can estimate the value of $x^\prime_F$ using the freeze-out approximation with $\Gamma_{3 \rightarrow 2}(x^\prime_F)=H(x_F=x^\prime_F/\xi_F)$, where $\xi_F=T_{\rm SM,F}/T_{\chi,{\rm F}}$ is the ratio of the two temperatures at the freeze-out point. This condition implies the following transcendental equation whose solution approximately determines $x^\prime_F$:
\begin{equation}
x^\prime_F + 2 \ln x^\prime_F = 27.94+ \frac{3}{4} \ln \left(\frac{(1 {~\rm GeV}) m_\chi^4 \langle \sigma v^2 \rangle_{3\chi \rightarrow 2\chi}}{\xi_i^2}\right) - \frac{3}{8} \ln g_*(x_F).
\label{Eq:freeze_32}
\end{equation}

\subsection{Big-bang nucleosynthesis constraints on $N_{\nu}$}
The effective number of relativistic degrees of freedom in the big-bang nucleosynthesis epoch determines the Hubble expansion rate during BBN, and thereby controls the freeze-out dynamics of the weak-interaction processes governing the primordial abundance of different chemical elements. A detailed analysis of these reactions, and its comparison with recent data on primordial abundance of different nuclei can thus be used to obtain an upper bound on the number of relativistic degrees of freedom in the BBN epoch. 

Therefore, if the cannibal DM species is relativistic during BBN, with $T_\chi > m_\chi$ at temperatures of around $T_{\rm BBN} \sim 1$ MeV, it will contribute to the Hubble expansion, leading to significant constraints. Although the BBN reactions take place within some range of the photon temperature, for our order-of-magnitude estimate we take the BBN temperature to be $T_{\rm BBN} \sim 1$ MeV. At this temperature, the relative heating of the photon bath due to electron-positron annihilations has not yet occurred, and therefore the neutrino temperature and the photon temperature remain the same. With this, the energy density contributed by a relativistic $\chi$ species at $T_{\rm BBN}$ is given by
\begin{equation}
\rho_\chi = \frac{\pi^2 g_\chi}{30} \frac{T_{\rm BBN}^4}{\xi_i^4},
\end{equation}
where, as defined earlier, the temperature ratio $\xi$ remains fixed at its initial value 
$\xi_i = \left(T_{\rm SM}/T_\chi\right)_i$, for 
 $T_\chi \geq m_\chi$. Thus the contribution of $\chi$ particles to the effective number 
of neutrinos at BBN is given by
\begin{equation}
\Delta N_\nu = \frac{4}{7} \frac{g_\chi}{\xi_i^4}.
\label{eq:BBNNnu}
\end{equation}

Using the recent analysis by Fields {\it et al.}~\cite{Fields:2019pfx}, we take the BBN constraint on the effective number of neutrinos to be
\begin{equation}
N_\nu = 2.878 \pm 0.278, ~~~~~~ 68\%{\rm ~C.L. ~Limit}.
\end{equation}
Thus, the $2\sigma$ constraint on $\Delta N_\nu$, taking the SM value of $N_\nu=3$ translates to a lower bound on $\xi_i$ if the DM is relativistic at the BBN epoch as
\begin{equation}
\xi_i > 1.07, ~~~~~~~~~95\%{\rm ~C.L. ~Constraint ~from ~BBN, ~for} ~~m_\chi \xi_i < T_{\rm BBN},
\end{equation}
where, we have taken $g_\chi=1$. This roughly implies that for a DM mass less than an MeV, its temperature at the BBN epoch should be lower than the SM temperature. In the scenario under study, the BBN constraint is independent of the DM couplings. This is because in order to be constrained by BBN, the DM needs to be relativistic at the BBN epoch, and since we are focussing on the cannibal process freezing-out when the DM is non-relativistic, the cannibal process always takes place after BBN for such DM mass and $\xi_i$ values. The BBN constraint is shown as the light green shaded region in Fig.~\ref{Fig:Cosmo_constraint1}. Our estimate for the BBN constraint is an approximate one, while a more detailed analysis of the effect of cannibal species on light element abundances require significant numerical computations~\cite{Kawano:1992ua,Arbey:2011nf}, which are beyond the scope of the present study.

\subsection{Constraints from the cosmic microwave background power spectrum}
The strongest constraint on the cannibal DM parameter space comes from its impact on the cosmic microwave background power spectrum. As mentioned in the introduction, this impact is largest if the $3\chi \rightarrow 2\chi$ process freezes-out during the matter-dominated epoch~\cite{Carlson:1992fn, Buen-Abad:2018mas, Heimersheim:2020aoc}. In contrast, in this study, we are focussing on a scenario in which the cannibal process freezes out during the radiation dominated epoch. Even in this case, as we shall see in the following, the CMB constraints turn out to be the most significant one. Essentially, since during the cannibal phase the DM temperature falls only logarithmically with the scale factor, it may end up being warmer than standard cold DM during the CMB epoch. Thus, for a given DM mass, the initial DM temperature needs to be accordingly colder to begin with -- smaller the mass, stronger the requirement. Detailed studies of the growth of matter density perturbations with cannibal DM have been performed in recent years, along with their impact on the CMB anisotropies~\cite{Buen-Abad:2018mas, Heimersheim:2020aoc}. As shown in Ref.~\cite{Heimersheim:2020aoc}, CMB data requires the cannibal DM to remain sufficiently non-relativistic at the time of photon decoupling so that it behaves like a cold DM around this epoch. The current CMB constraints can be approximately translated as a constraint on the DM temperature to mass ratio at the photon decoupling epoch with scale factor $a_{\rm LS}$ as~\cite{Heimersheim:2020aoc}
\begin{equation}
\frac{T_\chi (a_{\rm LS})}{m_\chi} < 10^{-5}, ~~~~~~~~{\rm at}~T_{\rm SM} (a_{\rm LS})\sim 0.26 {~\rm eV},
\label{Eq:CMB_1}
\end{equation}
where $a_{\rm LS} =a_0 T_0/T_{\rm SM} (a_{\rm LS}) \simeq 9 \times 10^{-4}$, 
$T_0$ being the present CMB photon temperature, with the present scale factor $a_0 = 1$.

In order to compute $T_\chi (a_{\rm LS})$, we note that after the freeze-out of the cannibal process at $a_{\rm FO}$, the DM temperature scales as $1/a^2$. We also note that the temperature ratio $\xi_{\rm FO}$ at the epoch of freeze-out of the cannibal process can be obtained using the conservation of the co-moving entropy in the DM as follows:
\begin{equation}
\xi_F = \xi_i {x^\prime_F}^{5/6} e^{(1-x^\prime_F)/3},
\end{equation}
where, $x^\prime_F =m_\chi/T_\chi (a_{\rm FO})$, details of the derivation being shown in Appendix~\ref{AppendixB1}. Combining these 
results, we obtain
\begin{equation}
\frac{T_\chi (a_{\rm LS})}{m_\chi} = \frac{T_{\rm SM} (a_{\rm LS})^2}{m_\chi^2} \xi_i^{-2} {x^\prime_F}^{-2/3} e^{2(x^\prime_F-1)/3}. 
\label{Eq:CMB_2}
\end{equation}
Therefore, once the freeze-out epoch for the cannibal process $x^\prime_F$ is obtained as a function of the DM mass $m_\chi$, $\xi_i$ and the DM self-couplings, we can use Eqs.~\ref{Eq:CMB_1} and~\ref{Eq:CMB_2} to determine the CMB constraints on the DM parameter space. The CMB constraints are shown by the light blue shaded region in Fig.~\ref{Fig:Cosmo_constraint1}. As we can see by comparing the regions disallowed by CMB for $\lambda=0.1$ and $\lambda=4\pi$, for a given DM mass, higher values of $\xi_i$ are ruled out by CMB for larger values of the quartic coupling. This is because, larger the coupling, longer the DM stays in chemical equilibrium during which its temperature falls only logarithmically with the scale factor. Thus starting from the same initial $\xi_i$ the temperature ratio at freeze-out $\xi_{\rm FO}$ is consequently larger, leading to a higher DM temperature at the epoch of photon decoupling as well. Therefore, if we decrease the DM self-couplings, the CMB constraints become consequently weaker. 
We see from Fig.~\ref{Fig:Cosmo_constraint1} that for a $1$ keV mass DM, the CMB constraints require 
$\xi_i \gtrsim 19$ with $\lambda=0.1$, while $\xi_i \gtrsim 63$ with $\lambda=4\pi$, and increasing the DM mass by an order of magnitude weakens the CMB bound by a similar order. Even with $\lambda=4\pi$, the CMB constraints are not found to be sensitive for DM masses higher than around an MeV, for $\xi_i>0.1$.

\subsection{S-matrix unitarity limits}
We now discuss an important theoretical limit coming from the consideration of S-matrix unitarity, which implies a model-independent upper bound on the total DM inelastic cross-section through the optical theorem~\cite{Griest:1989wd, Hui:2001wy, Bhatia:2020itt}. For the thermally averaged $3 \rightarrow 2$ $s-$wave annihilation rate, the unitarity upper bound is given by~\cite{Bhatia:2020itt}
\begin{equation}
\langle \sigma_{3 \rightarrow 2} v_{\rm rel}^{2} \rangle_{\rm max,~s-wave}=  \frac{8\sqrt{2} \left(\pi x^\prime \right)^{2}}{g_\chi m_\chi^{5}},
\label{Eq:3to2_sigma_max}
\end{equation}
where $x^\prime = m_\chi / T_\chi$, and the number of DM degrees of freedom in our scenario is $g_\chi=1$. Thus for a given value of $m_\chi$, the maximum possible annihilation rate is determined by Eq.~\ref{Eq:3to2_sigma_max}. With this cross-section and mass, there is a value of $\xi_i$ for which the relic density requirement of $\Omega_\chi h^2 = 0.12$ is satisfied. If we take a $\xi_i$ value lower than this, the SM temperature is also lower at the freeze-out point. Therefore, there is less time for the dilution of the DM number density due to the expansion of the Universe since freeze-out, leaving us with a higher DM density at the present epoch, which is disallowed by the observations. Since the unitarity upper bound on the cross-section is model independent, the unitarity bound, as shown in Fig.~\ref{Fig:Cosmo_constraint1} using the light orange shaded region, furnishes a constraint valid for all scenarios realizing a cannibal DM undergoing $3\chi \rightarrow 2\chi$ processes.  It is important to note that when the DM temperature is different from the SM one, the unitarity upper bound on the DM mass also gets modified accordingly, and is a function of $\xi_i$. The reaction rate obtained in our scenario, with perturbative couplings, is always significantly lower than the unitarity limit in Eq.~\ref{Eq:3to2_sigma_max}.

\subsection{Lower limits on the couplings}
\begin{figure} [htb!]
\begin{center} 
\includegraphics[scale=0.37]{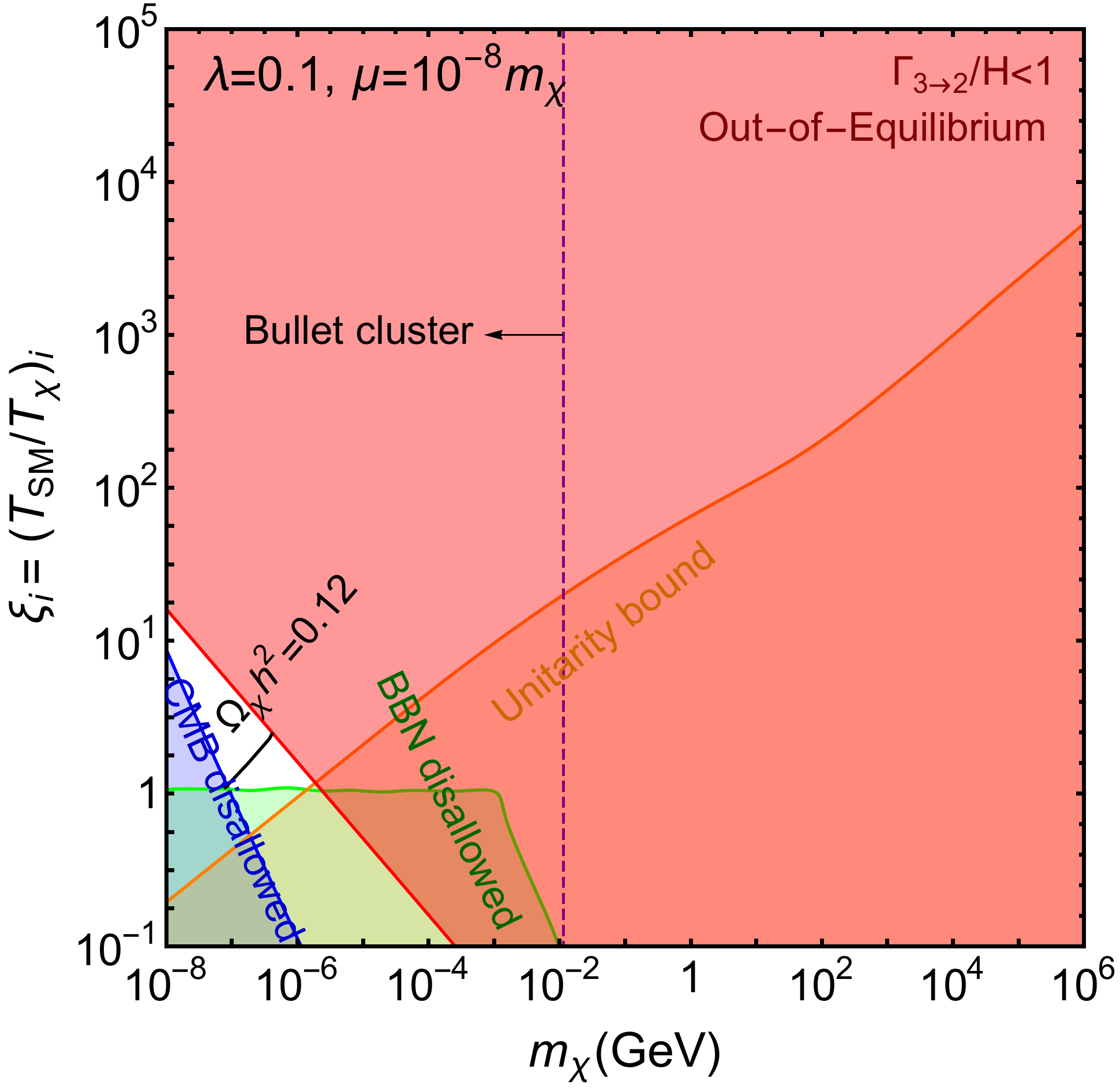} \hspace{0.0cm}
\includegraphics[scale=0.37]{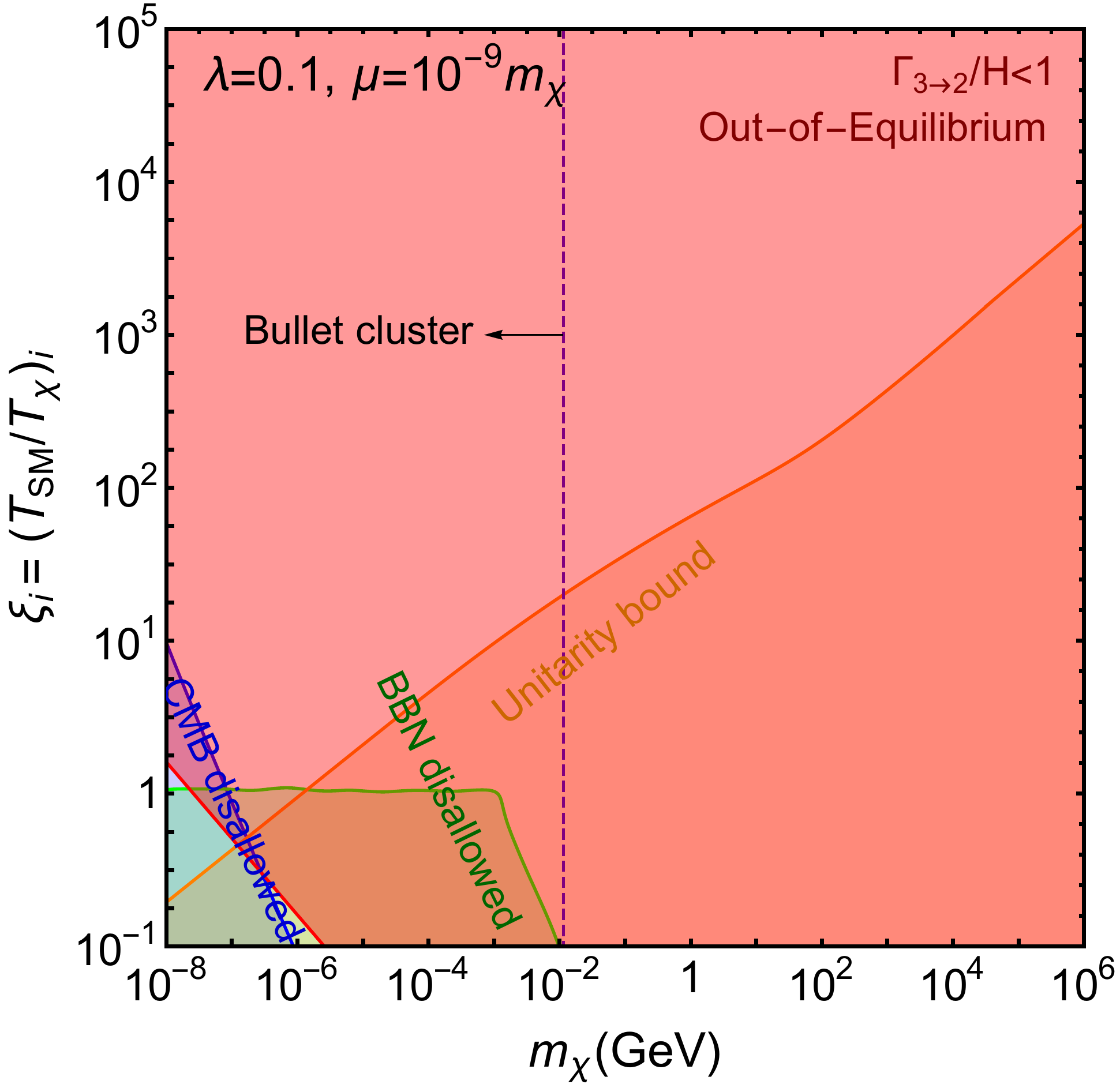}
\end{center}
\caption{\small{\it{Same as Fig.~\ref{Fig:Cosmo_constraint1}, for the 
DM self-couplings $\lambda=0.1, \mu=10^{-8}\,m_\chi$ (left panel), and $\lambda=0.1, \mu=10^{-9}\,m_\chi$ (right panel).} }}
\label{Fig:Cosmo_constraint2}
\end{figure}
As emphasized earlier, although the $m_\chi$ and $\xi_i$ parameter values for which the DM relic density is reproduced is less sensitive to coupling variations, the constraints, especially those from the requirement of chemical equilibrium of the $3\chi \rightarrow 2 \chi$ process and the CMB bounds are very sensitive to the couplings. 
Due to these constraints, we find that for a fixed value of the quartic coupling $\lambda$, one can obtain a lower bound on the trilinear coupling $\mu$, below which there is no allowed parameter space consistent with the relic density requirement, and the different cosmological constraints. In Fig.~\ref{Fig:Cosmo_constraint2} we  show the different constraints in the $m_\chi-\xi_i$ parameter space, for $\lambda = 0.1, \mu = 10^{-8}\,m_\chi$ (left panel) and 
$\lambda = 0.1, \mu = 10^{-9}\,m_\chi$ (right panel). As we can see from this figure, there exists a small parameter space for $\lambda = 0.1, \mu = 10^{-8}\,m_\chi$, while for $\lambda = 0.1, \mu = 10^{-9}\,m_\chi$ no viable parameter space exists. Therefore, in order to obtain a viable scenario for cannibal DM, 
one requires $\mu \gtrsim 10^{-9} m_\chi$ with $\lambda = 0.1$, which is a rather weak bound. Similarly, we find that for $\lambda = 4\pi$, the corresponding lower bound is $\mu \gtrsim 10^{-11} m_\chi$.

\subsection{Lyman$-\alpha$ and Bullet Cluster constraints}
\label{sec:LyalphaBul}

\begin{figure} [htb!]
\begin{center} 
\includegraphics[scale=0.30]{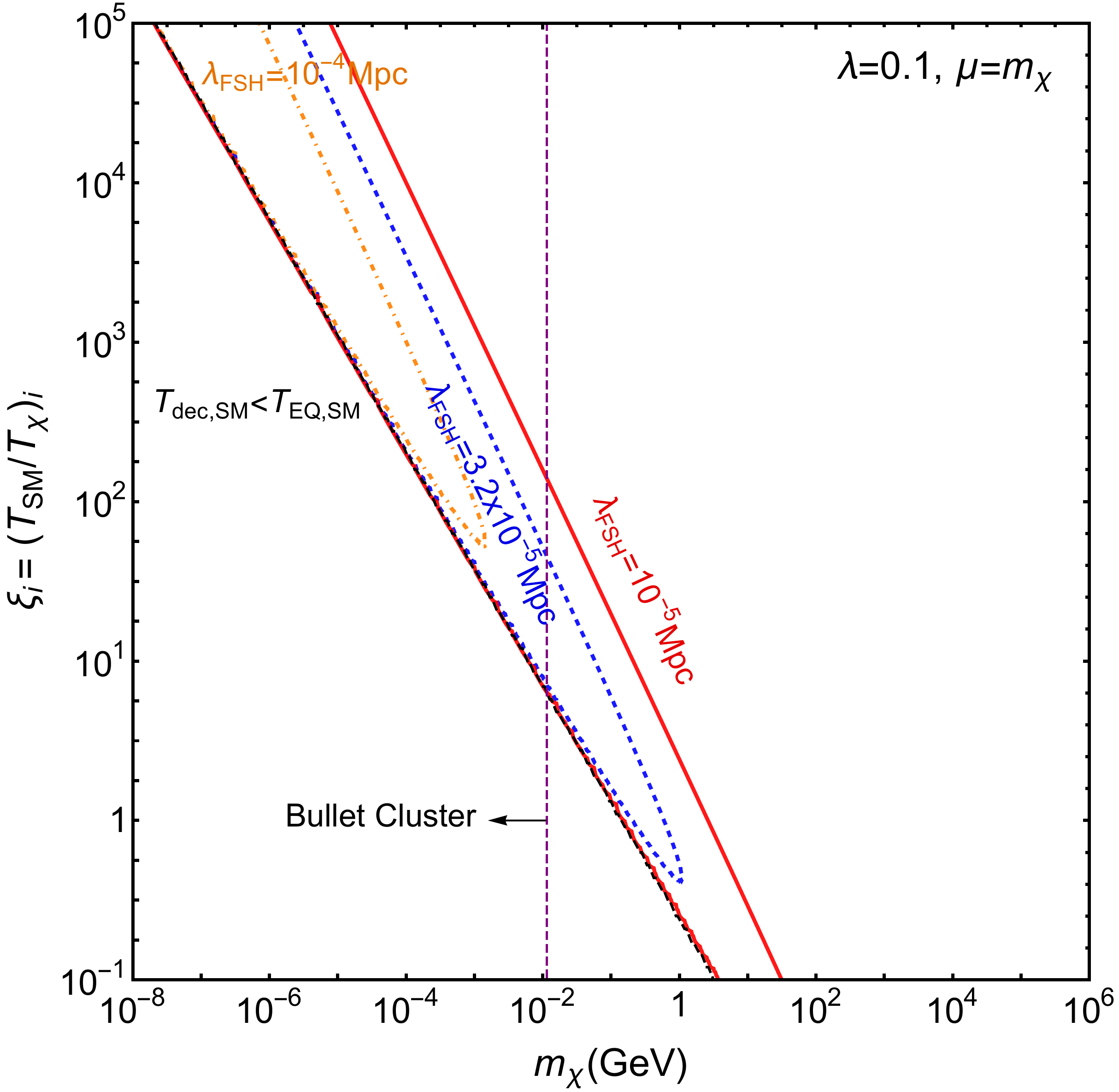} 
\includegraphics[scale=0.30]{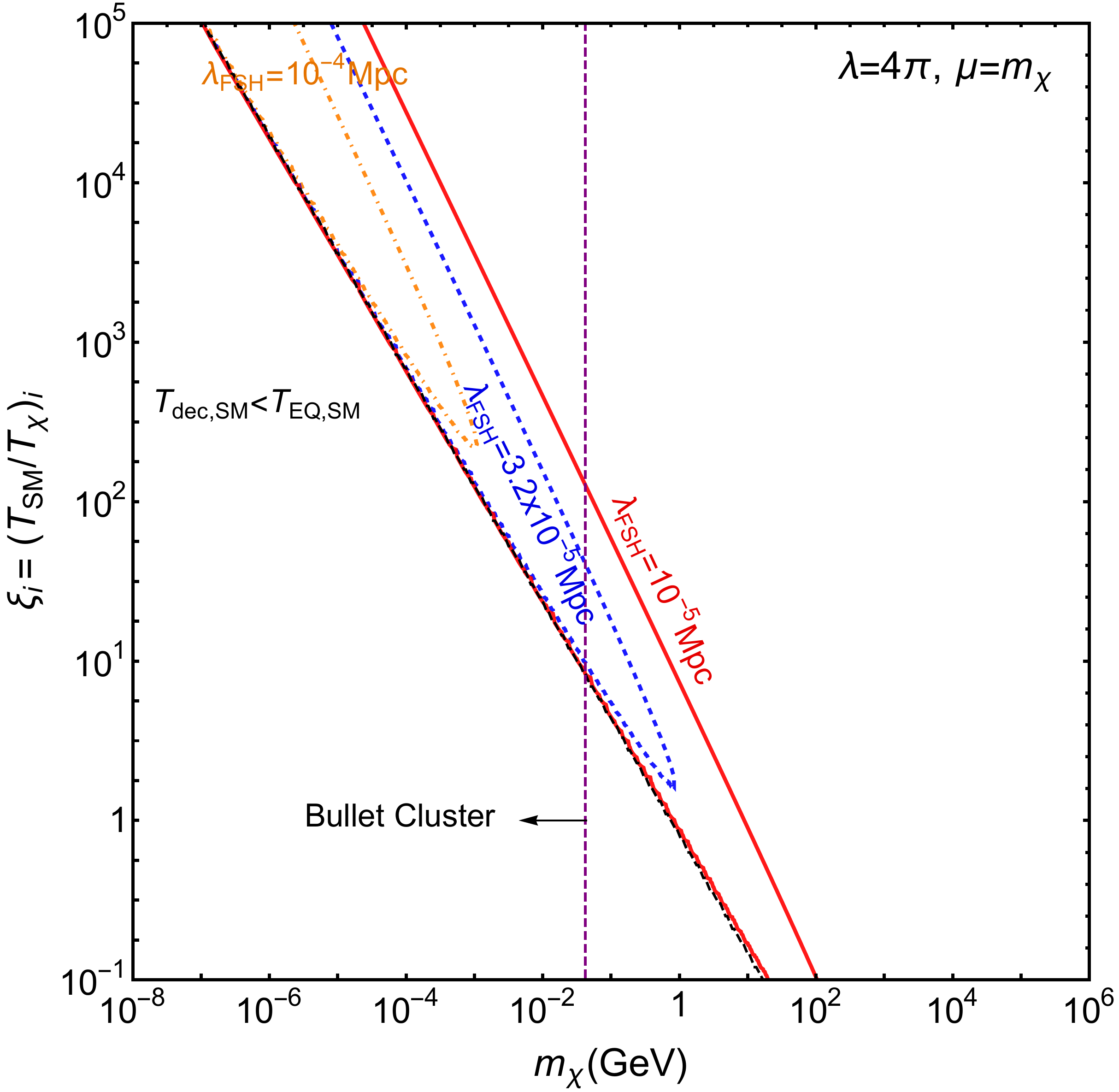}
\end{center}
\caption{\small{\it{Contours of fixed values of the free-streaming length $\lambda_{\rm FSH}$, with the DM quartic coupling $\lambda=0.1$ (left panel) and $\lambda=4\pi$ (right panel). The region in which the DM $2\chi \rightarrow 2\chi$ elastic scatterings decouple after matter-radiation equality, $T_{\rm dec, SM} < T_{\rm EQ, SM}$ is also shown. The Bullet Cluster limit on the elastic DM self-scattering, $\sigma_{2\chi \rightarrow 2\chi}/m_\chi < 1 {~\rm cm}^2/{\rm g}$, is indicated with the vertical dashed lines. See text for details.}}}
\label{Fig:Lyman_alpha}
\end{figure}
Since the Lyman$-\alpha$ measurements are sensitive to smaller scales for the matter power spectrum than the CMB, they may also lead to constraints on cannibal DM. As is well-known, collisionless particles can stream out of overdense regions into underdense regions until $t=t_{\rm EQ}$, $t_{\rm EQ}$ being the time of matter-radiation equality when the DM perturbations become Jeans unstable and begin to grow. As this process can lead to smoothing out of  inhomogeneities, it is constrained by the observed DM power spectrum~\cite{Kolb:1990vq}. To this end, we have made a simple estimate using the DM co-moving free-streaming length $\lambda_{\rm FSH}$, which approximately captures this effect. Here, $\lambda_{\rm FSH}$ is defined as the co-moving distance travelled by the DM particles from the time of decoupling of scattering reactions $t_{\rm dec}$ to the time of matter-radiation equality $t_{\rm EQ}$~\cite{Kolb:1990vq,Boyarsky:2008xj}:
\begin{eqnarray}
\lambda_{\rm FSH} = \int_{t_{\rm dec}}^{t_{\rm EQ}} \frac{\langle v(t) \rangle}{a(t)}dt.
\label{eqn:FSHlength}
\end{eqnarray}
This equation can also be re-written by changing variables from time $t$ to the scale factor $a(t)$ as
\begin{eqnarray}
\lambda_{\rm FSH} = \int_{a_{\rm dec}}^{a_{\rm EQ}} \frac{\langle v(a) \rangle}{a^2\,H(a)}da.
\label{eqn:FSHlength2}
\end{eqnarray} 
Performing this integral for the cannibal DM scenario under study, we obtain
\begin{equation}
\lambda_{\rm FSH} = \frac{3 M_{\rm Pl} T_0}{\sqrt{\rho_{R,0}}} \frac{\sqrt{x^\prime_F}}{\xi_F m_\chi} \ln \left(\frac{x_{\rm EQ}}{x_{\rm dec}}\right),
\label{Eq:FSH_canDM}
\end{equation}
where, $T_0 = 2.73$ K is the present CMB photon temperature, $\rho_{R,0}$ is the radiation energy density in the present epoch, $x_{\rm EQ}= m_\chi/T_{\rm EQ, SM}$ and $x_{\rm dec} = m_\chi / T_{\rm dec, SM}$, with $T_{\rm dec, SM}$ being the SM temperature at which the $2\chi \rightarrow 2\chi$ 
elastic scattering reactions decouple, and $T_{\rm EQ, SM}$ is the SM temperature at matter-radiation 
equality. Details of the derivation of Eq.~\ref{Eq:FSH_canDM} can be found in  
Appendix~\ref{AppendixB2}.
We have explicitly checked that in the parameter region of interest in which the 
$3\chi \rightarrow 2\chi$ process decouples when the DM is non-relativistic, the 
$2\chi \rightarrow 2\chi$ reaction 
always decouples after the $3\chi \rightarrow 2\chi$ reaction, i.e., $x_{\rm dec} > x_{\rm F}$. 

For the interaction Lagrangian given by Eq.~\ref{Eq:Lag}, the decoupling temperature for the $2\chi \rightarrow 2\chi$ elastic scattering process is governed by the following annihilation rate~\cite{Erickcek:2021fsu}
\begin{equation}
\langle \sigma v \rangle_{2\chi \rightarrow 2\chi}  \simeq (\sigma v)_{0,2\chi \rightarrow 2\chi} /  \sqrt{x^\prime},
\end{equation}
where,
\begin{equation}
(\sigma v)_{0,2\chi \rightarrow 2\chi} =  \left(\lambda - \frac{5\mu^2}{3 m_\chi^2}\right)^2 \frac{1}{64\,\pi^{3/2} m_\chi^2 }.\nonumber
\end{equation}

The decoupling temperature $x_{\rm dec}$ in Eq.~\ref{Eq:FSH_canDM}  can be approximately determined by the condition $n_\chi(x_{\rm dec}) \langle \sigma v \rangle_{2\chi \rightarrow 2\chi} (x_{\rm dec}) = H (x_{\rm dec})$, which leads to the following expression for $x_{\rm dec}$
\begin{equation}
x_{\rm dec} = \left(\frac{3}{\rho_{R,0}}\right)^{1/8} \frac{M_{\rm Pl}^{1/4} (\sigma v)_{0,2\chi \rightarrow 2\chi}^{1/2}}{\langle \sigma v^{2} \rangle_{3\chi \rightarrow 2\chi} ^{1/4}} \frac{T_0^{1/2}e^{(x^\prime_F-1)/2}}{\xi_i^{3/2} m_\chi^{1/2}}.
\end{equation}

We find that, in most of the parameter space of interest, in which the $3\chi \rightarrow 2\chi$ process decouples in the non-relativistic regime, the elastic scatterings decouple much later, often after the matter-radiation equality. Thus, for these latter parameter points, the free-streaming length as defined above vanishes. For the  region in which $T_{\rm dec, SM} > T_{\rm EQ, SM}$, we find the free-streaming length to be very small, of the order of  $10^{-4}$ or smaller with $\lambda=0.1$. We show in Fig.~\ref{Fig:Lyman_alpha} the contours of fixed values of $\lambda_{\rm FSH}$, with the quartic coupling $\lambda=0.1$ (left panel) and $\lambda=4\pi$ (right panel). We could not find any parameter points for which $\lambda_{\rm FSH} \gtrsim 10^{-4}$.The primary reason for such low values of the DM free-streaming is the late decoupling of the elastic $2\chi \rightarrow 2\chi$ scatterings. As we see from Fig.~\ref{Fig:Lyman_alpha}, for higher values of $\xi_i$, which leads to relatively earlier decoupling, $\lambda_{\rm FSH}$ is somewhat enhanced. For lower values of the DM quartic coupling the values of $\lambda_{\rm FSH}$ are correspondingly higher, for the same reason. Thus none of our parameter region of interest is constrained by the Lyman$-\alpha$ measurements which exclude $\lambda_{\rm FSH} \gtrsim 10^{-1}$~\cite{Irsic:2017ixq}~\footnote{For scenarios in which the elastic $2\chi \rightarrow 2\chi$ scatterings decouple late, considerations of dark acoustic oscillations can be relevant, and may lead to different constraints from the Lyman-$\alpha$ data than indicated by the free-streaming length analysis~\cite{Egana-Ugrinovic:2021gnu, Yunis:2021fgz, Garani:2022yzj}. Inclusion of these effects can lead to a lower bound of a few keV on the mass of DM that undergoes elastic self-scatterings. For these mass values, in our scenario the cannibal DM only constitutes about $1\%$ of the total DM density, for couplings $\mu=m_\chi$ and $\lambda=4\pi$ or $\lambda=0.1$ (see, Figs.~\ref{Fig:Cosmo_constraint1} and~\ref{Fig:DM_fraction}) and therefore, the constraints will be significantly altered compared to the studies in Refs.~\cite{Egana-Ugrinovic:2021gnu, Yunis:2021fgz, Garani:2022yzj}, the details of which is beyond the scope of this work.}. 

DM elastic self-scattering cross-sections, which played a major role in determining the DM free-streaming in the discussion above, can be constrained using the observations of colliding galaxy clusters, such as the Bullet Cluster~\cite{Randall:2008ppe, Robertson:2016xjh}, which indicate an approximate upper bound of $\sigma_{2\chi \rightarrow 2\chi}/m_\chi < 1 {~\rm cm}^2/{\rm g}$. This bound is shown with the vertical dashed lines in Fig.~\ref{Fig:Lyman_alpha}, as well as in Figs.~\ref{Fig:Cosmo_constraint1} and \ref{Fig:Cosmo_constraint2}. We find that with $\mu=m_\chi$, the Bullet Cluster observations disfavour a DM mass of $m_\chi \lesssim 40$ MeV for $\lambda=4\pi$, and $m_\chi \lesssim 10$ MeV for $\lambda=0.1$, for the scenario considered in this study, where the limit is subject to astrophysical uncertainties. However, the relationship between the $3\chi \rightarrow 2\chi$ and $2\chi \rightarrow 2\chi$ scattering rates is model dependent, and in the context of the cannibal DM cosmology discussed in the previous sections, a $2\chi \rightarrow 2\chi$ rate necessary to maintain kinetic equilibrium for the DM until the freeze-out of the cannibal process suffices.

\section{Summary}
\label{sec:sec4}
To summarize, we have studied a scenario in which DM has only gravitational interaction with the standard model (SM) particles at low temperatures. Such a DM may be internally thermalized, and may undergo number-changing self-scatterings in the early Universe, eventually freezing out to produce the observed DM abundance. If the number-changing reactions, such as a $3 \rightarrow 2$ process, take place when the DM is non-relativistic, DM cannibalizes itself to cool much slower than standard non-relativistic matter during the cannibal phase. It has been shown in earlier studies that if the cannibal phase takes place during the matter-dominated epoch, there are very strong constraints from structure formation. We considered instead scenarios in which the cannibal freeze-out happens during 
the radiation-dominated epoch, and showed that cannibal DM decoupled from the SM can be a viable possibility, consistent with all present cosmological constraints.

In order to accurately determine the abundance of the cannibal DM, we solved the coupled evolution equations of the DM temperature and density. We observed that the relic density evolves very slowly as a function of the DM self-couplings, and a factor of $125$ increase in the DM quartic coupling only leads to a factor of \sout{4.5} 
{\color{red} 1.6} decrease in the abundance. On the other hand, for the same variation 
of the quartic coupling, the DM temperature at late epochs may vary by upto one order of magnitude, thereby affecting the resulting cosmological constraints 
significantly. We also found that for an accurate modelling of the DM temperature, solving the 
coupled evolution equations is necessary, as the analytic approximations using co-moving DM entropy 
conservation can result in upto an order of magnitude error in the value of the DM temperature at 
late epochs.

We evaluated the constraints on the DM mass and initial temperature from the cosmic-microwave background power spectrum, the big-bang nucleosynthesis limits on the relativistic degrees of freedom, the Lyman-$\alpha$ limits on the DM free-streaming length and the theoretical upper bound on the $3 \rightarrow 2$ annihilation rate from $S-$matrix unitarity. The $95\%{\rm ~C.L.}$ BBN bounds require the initial temperature ratio $\xi_i = \left(T_{\rm SM}/T_\chi\right)_i > 1.07$, for DM particles that are relativistic during BBN. This roughly implies that for a DM mass less than an MeV, its temperature at the BBN epoch should be lower than the SM temperature. The unitarity bound sets a model independent lower limit on $\xi_i$ for a given DM mass, and is applicable to all cannibal DM scenarios undergoing $3 \rightarrow 2$ annihilations. None of our parameter region of interest is constrained by the Lyman$-\alpha$ measurements, since the DM free-streaming length is found to be rather small. This is primarily because of the very late decoupling of the elastic $2 \rightarrow 2$ DM self-scatterings, which, in a large region of the parameter space, takes place after the matter-radiation equality. However, scenarios leading to a large DM elastic self-scattering may be constrained further by Bullet Cluster observations.

The CMB matter power spectrum leads to  the strongest cosmological constraint on cannibal DM, as depending upon its initial temperature, cannibal DM might not be sufficiently cold during the photon decoupling epoch, leading to modifications in the power spectrum. For larger values of DM self-couplings, the CMB constraints become stronger. 
For a $1$ keV mass DM, the CMB constraints require $\xi_i \gtrsim 19$ with $\lambda=0.1$, while 
$\xi_i \gtrsim 63$ with $\lambda=4\pi$, and increasing the DM mass by an order of magnitude weakens the CMB bound by a similar order. Even with $\lambda=4\pi$, the CMB constraints are not found to be sensitive for DM masses higher than around an MeV, for $\xi_i>0.1$. 

Although the CMB bounds make larger values of $\xi_i$ preferred, the requirement of chemical 
equilibrium when the DM is non-relativistic puts an upper bound on the possible values of $\xi_i$ for 
a given DM mass. Thus, combining both these requirements leads to a significant restriction on the DM 
parameter space, although large regions remain viable.  We find that for the DM quartic coupling 
$\lambda=4\pi$, and trilinear coupling $\mu = m_\chi$, a scalar cannibal DM with mass in the range of 
around 28.2 keV to 707 TeV can make up the observed DM density 
and satisfy all the constraints, when the initial temperature ratio is in the range 
$3.3 \lesssim \xi_i \lesssim 9120$. 
For $\lambda=0.1$, the allowed mass is in the range of 9.3 keV to $5.7$ TeV, with 
$ 2.7 \lesssim \xi_i \lesssim 1950$. 
We find that for couplings as small as $\mu=10^{-8}m_\chi$ and $\lambda = 0.1$, 
some of the parameter space are viable, with the allowed DM mass being in the narrow range of 
79 eV to 407 eV, with $ 1.1 \lesssim \xi_i \lesssim 2.45$.  
We further observe that with $\lambda=0.1$, for $\mu \lesssim 10^{-9} m_\chi$ 
no parameter space remains viable for a cannibal DM candidate consistent with all the cosmological constraints. Similar conclusions are obtained for  
$\mu \lesssim 10^{-11} m_\chi$ with $\lambda=4\pi$. 

To conclude, a cannibal DM freezing out during radiation domination is very much allowed by the structure formation constraints, even when the DM cannot dissipate the heat generated during the cannibal phase to the SM sector, as long as it is sufficiently cold compared to the SM to begin with, but not so cold that chemical equilibrium of the $3 \rightarrow 2$ process is not achieved. Future more precise determination of the matter power spectrum remains the best probe for the allowed scenarios of such DM.

\appendix
\section{Appendix}
\label{Appendix}

\subsection{Some useful formulae}
\label{AppendixA}
\begin{figure} [htb!]
\begin{center} 
\includegraphics[scale=0.8]{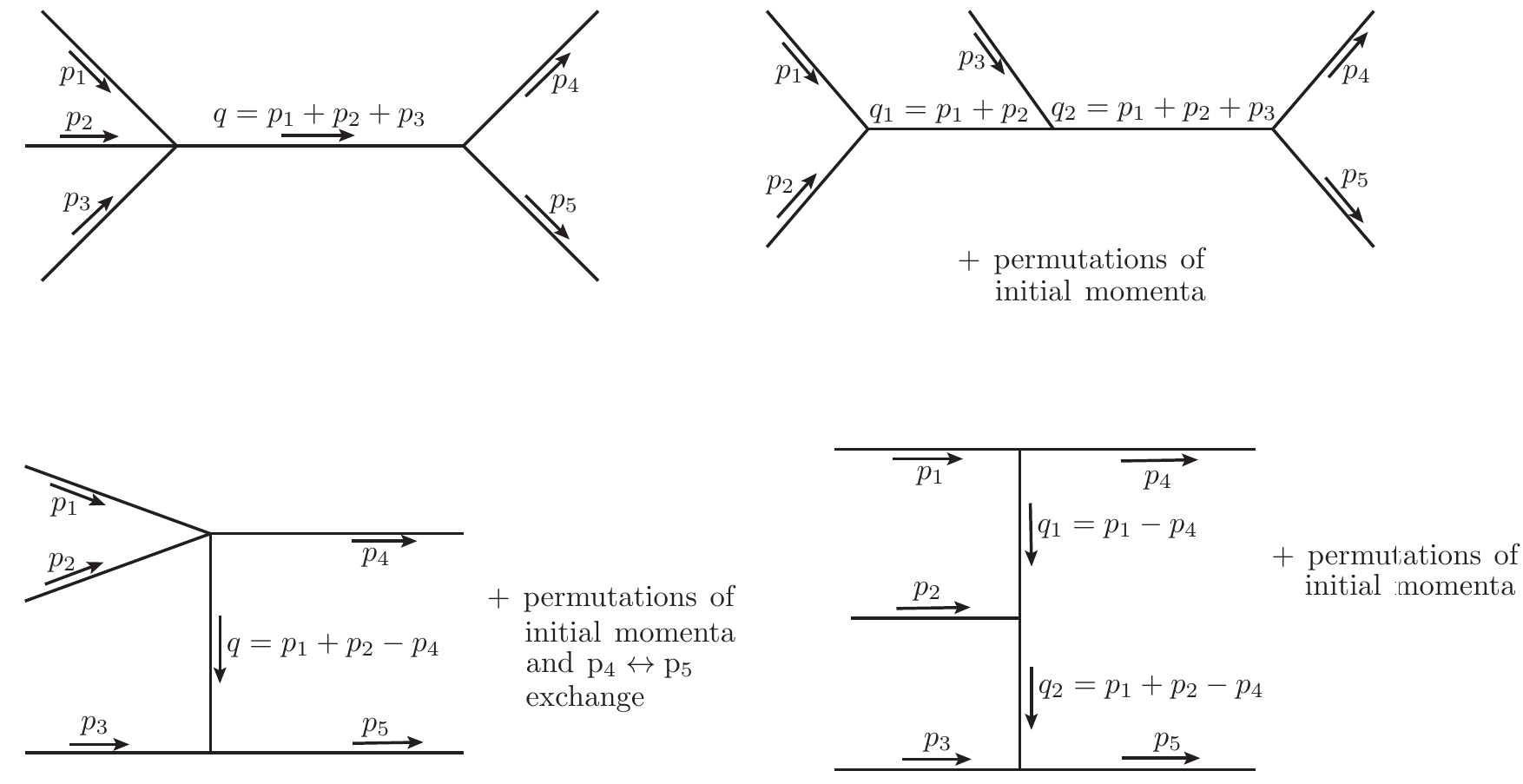}\\
\vspace{5mm}
\includegraphics[scale=0.8]{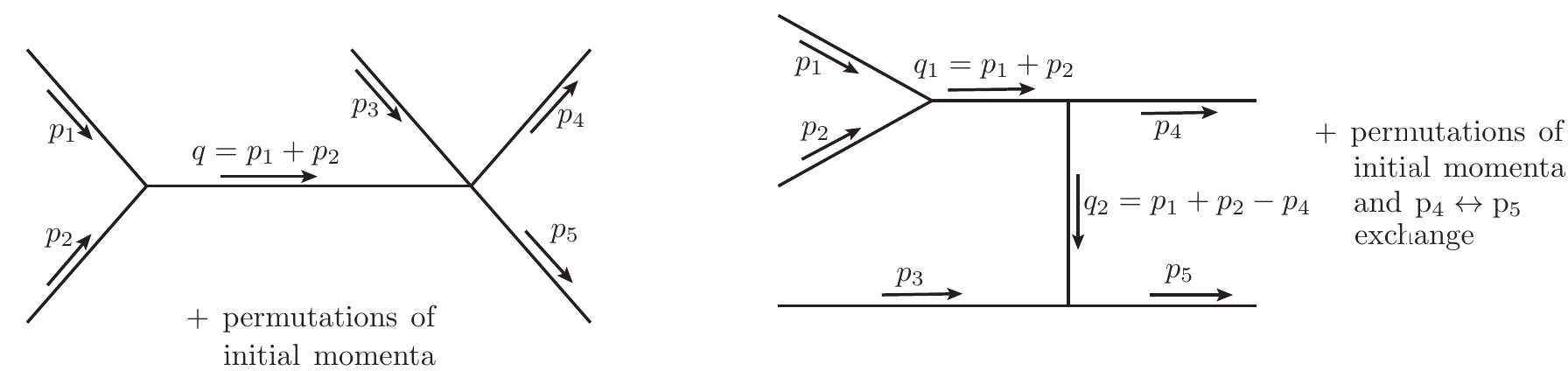} 
\end{center}
\caption{\small{\it{Different class of Feynman diagrams contributing to the $3\chi \rightarrow 2\chi$ number changing process, with the additional diagrams within each class obtained by the permutations of different momenta, as indicated.}}}
\label{Fig:Feyn_diag}
\end{figure}

With the interaction Lagrangian density given by Eq.~\ref{Eq:Lag}, the different class of Feynman diagrams which contribute 
to the $3\chi \rightarrow 2\chi$ number changing process are shown in  
Fig.~\ref{Fig:Feyn_diag}, with the additional diagrams within each class obtained by the permutations of different momenta, as indicated in this figure. The corresponding matrix element is given by the coherent sum of these diagrams and is found to be

\begin{equation}
\mathcal{M}_{3\chi \rightarrow 2\chi} = \frac{5\,\mu}{8\,m^2_\chi}\left(3\lambda - \frac{\mu^2}{m^2_\chi} \right).
\label{Eq:3to2_mat_elem}
\end{equation}

Using this matrix element, we compute the annihilation rate $\langle \sigma v^2 \rangle_{3\chi\rightarrow2\chi}$ defined in Eq.~\ref{Eq:sigmav2} and $\sigma v^2$ 
defined in Eq.~\ref{Eq:sigmav2_noavg}.

The thermally averaged $3\chi \rightarrow 2\chi$ cross-section is given by,
\begin{align}
\label{Eq:C0_Integral}
\langle \sigma v^2 \rangle_{3\chi\rightarrow2\chi} = \frac{1}{3!2!}\frac{1}{(n_{\chi}^{\rm eq})^3} \int d\Pi_1 d\Pi_2 d\Pi_3 d\Pi_4 d\Pi_5\,(2\pi)^4\,\delta^4(p_{\chi_1}+p_{\chi_2}+p_{\chi_3}-p_{\chi_4}-p_{\chi_5})\,|\mathcal{M}|^2_{3\chi \rightarrow 2\chi},
\end{align} 

When $|\mathcal{M}|^2_{3\chi \rightarrow 2\chi}$ is constant the phase space integrals 
in Eq.~\ref{Eq:C0_Integral} can be done analytically, and one obtains~\cite{Kuflik:2017iqs}, 
\begin{equation}
\langle \sigma v^2 \rangle_{3\chi\rightarrow2\chi} = \frac{\sqrt{5}}{2304\pi\,m^3_\chi}\,|\mathcal{M}|^2_{3\chi \rightarrow 2\chi}.
\end{equation}

Therefore, for our model the thermally averaged 
$\langle \sigma v^2 \rangle_{3\chi \rightarrow 2\chi} $ is~\cite{Erickcek:2021fsu}

\begin{equation}
\langle \sigma v^2 \rangle_{3\chi \rightarrow 2\chi} \simeq \frac{25\sqrt{5} \mu^2}{147456 \pi m_\chi^7} \left(3\lambda - \frac{\mu^2}{m^2_\chi} \right)^2. 
\end{equation}

The collision integral in Eq.~\ref{Eq:C2_reduced} can be evaluated in a similar way, by expressing the $\frac{|\textbf{p}_i|^2}{E_i}$ factors in terms of the integration variables $s, m_{12}, m_{23}$ and $E_0$. The resulting expressions are tabulated below.
\begin{equation}
\begin{split}
&\frac{|\textbf{p}_1|^2}{E_1} = \frac{(s-m_{23}^2+m_{\chi}^2)^2-4E_0^2m_{\chi}^2}{2E_0(s-m_{23}^2+m_{\chi}^2)}\\
&\frac{|\textbf{p}_2|^2}{E_2} = \frac{2E_0^2-2s+m_{12}^2+m_{23}^2-2m_{\chi}^2}{2E_0}-\frac{2E_0m_{\chi}^2}{2E_0^2-2s+m_{12}^2+m_{23}^2-2m_{\chi}^2}\\
&\frac{|\textbf{p}_3|^2}{E_3} = \frac{(s-m_{12}^2+m_{\chi}^2)^2-4E_0^2m_{\chi}^2}{2E_0(s-m_{12}^2+m_{\chi}^2)}\\
&\frac{|\textbf{p}_4|^2}{E_4} = \frac{s^2-4E_0^2m_{\chi}^2}{2E_0s}\\
&\frac{|\textbf{p}_5|^2}{E_5} = \frac{(2E_0^2-s)^2-4E_0^2m_{\chi}^2}{2E_0(2E_0^2-s)}.
\end{split}
\end{equation}

\subsection{Derivation of $\xi_F$ and $\lambda_{\rm FSH}$}
\label{AppendixB}

\subsubsection{Derivation of $\xi_F$}
\label{AppendixB1}
Using co-moving entropy conservation for dark matter during the cannibal phase which starts at 
$T_\chi (a_i) = m_\chi$ and ends at $T_\chi(a_{\rm FO}) = m_\chi/x^\prime_F$, we obtain
\begin{eqnarray}
\frac{a_{\rm FO}}{a_{i}} &=& x^{\prime\,1/6}_F\,e^{(x^{\prime}_F-1)/3}, 
\label{eq:xiFOeq1}
\end{eqnarray}
where cannibal co-moving entropy density is defined as 
$s (T_\chi) = m_\chi n_\chi (T_\chi)/T_{\chi}$.  
Now, the temperature ratio at freeze-out is related to the corresponding ratio at 
$T_\chi = m_\chi$ as follows: 
\begin{eqnarray}
\xi_F/\xi_i &=& \frac{T_{\rm SM} (a_{\rm FO})}{T_\chi (a_{\rm FO})}\,\frac{T_\chi (a_i)}{T_{\rm SM} (a_i)} = \frac{a_i}{a_{\rm FO}}\,x^\prime_F,\\
\Rightarrow \xi_F &=& \xi_i\,x^{\prime\,5/6}_F\,e^{(1-x^{\prime}_F)/3},
\end{eqnarray}
where we have used $a_i T_{\rm SM} (a_i)= a_{\rm FO} T_{\rm SM} (a_{\rm FO})$ 
and Eq.~\ref{eq:xiFOeq1}.

\subsubsection{Derivation of $\lambda_{\rm FSH}$}
\label{AppendixB2}
The free-streaming length is given by:
\begin{eqnarray}
\lambda_{\rm FSH} = \int_{a_{\rm dec}}^{a_{\rm EQ}} \frac{\langle v(a) \rangle}{a^2\,H(a)}da,
\label{eqn:FSHlength2A}
\end{eqnarray} 
where, $a = T_0/T_{\rm SM} = \frac{T_0}{m_\chi}x$. Now, we can write, 
\begin{equation}
\frac{da}{a^2 H(a)} = \frac{\sqrt{3}\,M_{\rm Pl}\,T_0}{\sqrt{\rho_{R,0}}\,m_\chi}\,dx, 
\label{eqn:FSHlength2B}
\end{equation}
and $\langle v(a) \rangle = \sqrt{\dfrac{3 T_\chi}{m_\chi}} = \dfrac{\sqrt{3 x^\prime_F}}{\xi_F\,x}$.

Therefore, we can write,
\begin{eqnarray}
\lambda_{\rm FSH} &=& \frac{3\,M_{\rm Pl}T_0}{\sqrt{\rho_{R,0}}}\frac{\sqrt{x^\prime_F}}{m_\chi\,\xi_F}\,\int_{x_{\rm dec}}^{x_{\rm EQ}} \frac{dx}{x},\\
\Rightarrow \lambda_{\rm FSH} &=& \frac{3 M_{\rm Pl} T_0}{\sqrt{\rho_{R,0}}} \frac{\sqrt{x^\prime_F}}{\xi_F m_\chi} \ln \left(\frac{x_{\rm EQ}}{x_{\rm dec}}\right).
\label{eqn:FSHlength2C}
\end{eqnarray}

\section*{Acknowledgment}
We thank Disha Bhatia for helpful discussions and collaboration in the early stages of this project. We also thank Sougata Ganguly, Dhiraj Kumar Hazra and Deep Ghosh for useful discussions, and Utpal Chattopadhyay for computational help.

\end{document}